\documentclass[12pt]{article}
\usepackage{graphicx}
\usepackage{amssymb}
\usepackage{amsmath}
\usepackage{cite}
\usepackage{hyperref}
\usepackage{bbold}
\usepackage{multirow}
\usepackage{tikz}

\numberwithin{equation}{section}
\numberwithin{table}{section}

\def\beq{\begin{equation}}
\def\eeq{\end{equation}}
\def\be{\begin{equation}}
\def\ee{\end{equation}}
\def\bea{\begin{eqnarray}}
\def\eea{\end{eqnarray}}

\def\Re{{\rm Re\,}}
\def\Im{{\rm Im\,}}

\DeclareRobustCommand{\SkipTocEntry}[4]{}
\RequirePackage{color}

\newcommand{\cM}{\mathcal{M}}
\newcommand{\cN}{\mathcal{N}}

\newcommand{\bbZ}{\mathbb{Z}}

\newcommand{\id}{{\mathbb1}}

\DeclareMathOperator{\sign}{sign}

\begin{document}

\begin{titlepage}
\begin{center}
\rightline{\small }


\vskip 2cm

{\Large \bf An index for flux vacua}
\vskip 1.2cm

Severin L\"ust

\vskip 0.5cm

{\it Laboratoire Charles Coulomb (L2C), Université de Montpellier, CNRS,\\ F-34095, Montpellier, France}

\vskip 0.8cm

{\tt severin.lust@umontpellier.fr} 

\end{center}

\vskip 1cm

\begin{center}  {\bf Abstract }\\ \end{center}

\noindent
We propose to use the winding number of the gradient of a scalar potential as  a simple topological index that relates critical points in the interior of the scalar field space to the behavior of the potential at the (asymptotic) boundary of the field space.
We demonstrate this technique for supersymmetric flux compactifications of M-theory on Calabi-Yau four-folds, and use the Fermat sextic as a simple, one-parameter example.

\noindent

\vfill

\noindent
May 2024

\end{titlepage}




\section{Introduction}

One of the most important problems of string phenomenology is to decide whether a given compactification (for example specified by a choice of compactification space, fluxes, and branes) gives rise to a consistent, (meta)stable string theory vacuum.
While this question is often straightforward to answer for the case of unbroken, extended supersymmetry (i.e.,\ $\cN \geq 2$ in four dimensions), 
it becomes a much more difficult problem if only minimal ($\cN = 1$) supersymmetry is preserved or if supersymmetry is broken completely.

In a four-dimensional theory with unbroken $\cN = 2$ supersymmetry, supersymmetry is powerful enough to guarantee the absence of various quantum corrections.
In particular, such theories can have an exactly vanishing potential and hence allow for an exact moduli space of vacua.
Moreover, many non-trivial corrections to the field space metric can be computed exactly, for example, in the vector multiplet sector by means of mirror symmetry.%
\footnote{For quantum corrections to the hypermultiplet moduli space see, for example, \cite{Alexandrov:2011va}.}

On the other hand, for $\cN = 1$ supersymmetry, the situation 
is much more intricate.
Quantum corrections are abundant and usually not all explicitly known (for an overview see, e.g., \cite{Palti:2020qlc, Gao:2022uop}, and references therein).
Most importantly, the scalar potential itself is not protected from corrections and hence generically non-vanishing.%
\footnote{It was suggested in \cite{Palti:2020qlc} (see also \cite{Kashani-Poor:2013iac, Grimm:2014aha}) that all allowed corrections are generically present unless the theory originates from one with a higher amount of supersymmetry.}
Whereas the holomorphic superpotential, $W$, is only corrected non-perturbatively, the K\"ahler potential also receives perturbative corrections.
Together this opens the door for various possible control issues (as, more generally, observed early on by Dine and Seiberg \cite{Dine:1985he}) and makes a stability analysis contingent on the precise knowledge of all relevant corrections.
Similar (and likely worse) problems arise if supersymmetry is broken completely.

These issues can be avoided if a non-trivial scalar potential is generated already at the classical level, and if the classical potential is parametrically larger than any additional quantum corrections.
This could be achieved, for example, by non-trivial background fluxes.
However, in the important case of IIB flux-compactification on a Calabi-Yau orientifold, only the complex structure moduli are stabilized by classical fluxes.
The remaining K\"ahler moduli are tree-level massless and might only get stabilized once quantum 
corrections are taken into account, potentially bringing back the problems outlined above.

Flux compactifications of IIB attracted a lot of attention as they are the base for both the KKLT \cite{Kachru:2003aw} and the large volume (LVS) \cite{Balasubramanian:2005zx} compactification scenarios.
These setups are among the most promising proposals for well controlled and meta-stable de Sitter vacua in string theory (for a recent, alternative approach see, e.g., also \cite{DeLuca:2021pej}).
However, it is still much debated in the literature if they can actually be realized in a fully consistent way.
Significant progress in this direction has recently been made in \cite{Demirtas:2021nlu,Demirtas:2021ote},
where $\cN=1$ supersymmetric AdS vacua along the lines of KKLT were constructed highly explicitly, including a multitude of correction terms to the effective action.
At the same time, a possible consistency issue with such AdS vacua at the level of the holographic AdS/CFT duality has been pointed out \cite{Lust:2022lfc} (see also \cite{BenaLiLust}).
Moreover, not all correction terms are  known fully explicitly yet (see, e.g.,~\cite{Kim:2022jvv, Alexandrov:2022mmy, Kim:2023cbh, Kim:2023sfs, Kim:2023eut, Schreyer:2024pml} for recent progress),
and various possible control issues in both the KKLT scenario \cite{Bena:2018fqc, Carta:2019rhx, Blumenhagen:2019qcg, Gao:2020xqh, Lust:2022xoq, Blumenhagen:2022dbo, Hebecker:2022zme, Schreyer:2022len} as well as in the LVS \cite{Junghans:2022exo, Junghans:2022kxg, Gao:2022fdi, Gao:2022uop, ValeixoBento:2023nbv} have been identified.
Interestingly, many of these results relate parametric control to large D3 tadpole charges, which has been argued to stand in possible conflict with complex structure moduli stabilization in \cite{Bena:2020xrh, Bena:2021wyr}.%
\footnote{For a more detailed account on the current literature on string theory de Sitter constructions and flux compactifications see, e.g., \cite{Danielsson:2018ztv, Cicoli:2023opf, VanRiet:2023pnx, McAllister:2023vgy}, and references therein.}

Non-withstanding whether parametrically controlled vacua of IIB orientifold compactifications can actually be found, a huge fraction of the moduli space and hence potentially also of the string theory landscape remains currently computationally inaccessible.
When moving towards the interior of the moduli space, quantum and curvature corrections become increasingly large, until eventually any series expansion of corrections around classical supergravity breaks down.
For example, when the volume of the compactification space approaches the string scale, curvature (and possibly other) corrections become large so that at some point a treatment in terms of classical geometry is not possible anymore.%
\footnote{For recent advances towards a better understanding of the interior of the moduli space of $\cN=1$ compactifications, see, e.g., \cite{Marchesano:2022avb, Wiesner:2022qys}.}

On the other hand, the asymptotic structure of the scalar field space and the scalar potential is significantly better understood than the interior of the moduli space.
For example, the size of many correction terms is typically controlled by the inverse volume of the compactification space. Hence, they become negligible in a decompactification limit where the volume is sent to infinity.

Asymptotic limits in scalar fields space have been investigated a lot and systematically in the context of the Swampland Program.
Here, it has been proposed under the name of the Swampland Distance Conjecture \cite{Ooguri:2006in} that any EFT with a consistent quantum gravity UV completion necessarily breaks down when approaching an infinite distance limit in scalar field space due to the appearance of an infinite tower of light states.
Also, a classification of possible infinite distance limits has been put forward, dubbed the Emergent String Conjecture \cite{Lee:2019wij}.
Moreover, it was conjectured \cite{Obied:2018sgi, Ooguri:2018wrx, Bedroya:2019snp} that any potential must satisfy a bound of the form $|\nabla V| \gtrsim V$ in such an asymptotic limit, implying the absence of any de Sitter vacua close to the infinite distance boundaries of field space.%
\footnote{Similar bounds on the so-called species scale and implications on effective potentials were discussed in \cite{vandeHeisteeg:2023uxj}, see also \cite{Castellano:2023stg, Castellano:2023jjt}.}
Consequently, the question whether cosmological models of accelerated expansion can be realized in asymptotic limits in field space gained a lot of recent attention (see, e.g., \cite{Rudelius:2022gbz, Calderon-Infante:2022nxb, Marconnet:2022fmx, Shiu:2023nph, Shiu:2023fhb, Hebecker:2023qke, VanRiet:2023cca, Andriot:2023wvg}).

In this note we want to explore the idea whether any information on critical points or minima of the potential in the interior of the field space can be inferred from its asymptotic behavior at the infinite distance boundaries of the field space.
For a scalar potential $V$ formally defined on an $n$-dimensional scalar manifold $\cM$ with (asymptotic) boundary $\partial \cM$, 
we suggest to use the 
winding number
of the gradient $\nabla V$ (understood as a vector field on $\cM$) as a topological index. 
As we review in the next section, there exists a suitably defined integral $I$ over the boundary $\partial \cM$,
\begin{equation}
    I = \int\limits_{\partial \cM}  \omega \left(\nabla V \right) \,, 
\end{equation}
where $\omega \left(\nabla V \right)$ is an $(n-1)$-form that depends only on the gradient of the potential $V$. 
The integral $I$ computes the higher dimensional analog of the winding number of the vector field $\nabla V$ on the boundary $\partial \cM$ of the scalar field space.
It is therefore an integer number and is equal to the sum of the index or winding number of $\nabla V$ at all isolated critical points and singularities of $V$.
Assuming hence that $V$ is actually non-singular,
this integral carries precisely the information we are after.

A few possible caveats can be anticipated:
1) Not all critical points of the potential are necessarily actual minima.
2) There might not exist a non-singular potential that is globally well-defined on the whole moduli space $\cM$.
3) The boundary $\partial \cM$ and the behavior of the potential on it might not be fully known.
At least the first issue can be circumvented by restricting to supersymmetric vacua of an $\cN=1$ F-term potential.
Here, every critical point $D_i W = 0$ of the K\"ahler-covariant derivative of the superpotential $W$ is guaranteed to correspond to an actual supersymmetric minimum of the potential $V \sim e^K (g^{i \bar \jmath} D_i W \overline D_{\bar \jmath} \overline W - |W|^2)$.
Hence, if we manage to compute the integral $I$ directly for the vector field $DW$ instead of $\nabla V$, it will encode information on the F-term minima of the potential, and not just on its critical points.

In the remainder of this paper we explore this proposal for classical flux vacua of M-theory on a Calabi-Yau four-fold \cite{Becker:1996gj, Dasgupta:1999ss}.
They are closely related to flux-vacua of type IIB string theory \cite{Giddings:2001yu,Grana:2001xn} and are characterized as the F-term minima with respect to the Gukov-Vafa-Witten superpotential \cite{Gukov:1999ya}.
Hence, we do not have to worry about the first issue in the above list.
Moreover, this setup has the benefit that not only the asymptotic structure of the superpotential $W$ but its exact form on the whole complex structure moduli space is in principle well known in terms of the underlying period integrals.
Furthermore, if we exclude certain special points and restrict to a single patch with respect to the infinite-order monodromies around these points, there indeed exists a well-defined, non-singular superpotential on the full moduli space.
Therefore, also the other two potential pitfalls are resolved here.
However, as we will see for the concrete example of the Fermat sextic later on, additional problems can arise 
in the form of branch cuts in complex structure moduli space.

The rest of this paper is organized as follows:
In Section~\ref{sec:index} we review some basic mathematical facts, and outline our general strategy in Section~\ref{sec:boundarypotentials}.
In Section~\ref{sec:fluxvacua} we present our main example: 
Flux-vacua of M-theory compactifactions on the Fermat sextic.
After a brief summary on M-theory flux compactifications, we summarize in Section \ref{sec:sextic} the complex structure moduli space geometry and period integrals of the sextic Calabi-Yau four-fold.
In Section~\ref{sec:sexticintegral} we discuss how to compute the index integral along the boundary of its moduli space and finally present a few examples in Section~\ref{sec:examples}.
In Section~\ref{sec:conclusions} we conclude and summarize our findings.
A few more technical results can be found in the appendix.

Two supplementary Mathematica notebooks are provided: one contains computations of the period integrals using the Frobenius method, and the other includes explicit computations and tools to illustrate the results.

\section{Preliminaries}\label{sec:preliminaries}

In this section we first summarize a few well known mathematical results. 
We then discuss how to use these results to infer information on the existence of critical points or minima of a scalar potential from its behavior at the boundary of the scalar field space.

\subsection{Index theory}\label{sec:index}

We start by discussing the real two-dimensional case.
Let $X = (X^1, X^2)$ be a differentiable vector field on an open subset  $U \subseteq \mathbb{R}^2$ and $C$ a closed curve (without self-intersections) in $U$.
Let us furthermore assume that $X$ has only a finite number of isolated zeros and singularities and that none of them lie on $C$.
We define the index $I_C$ by the contour integral
\begin{equation}\label{eq:contourindex}
    I_{C}(X) = \frac{1}{2 \pi} \oint_C \frac{\epsilon_{ij} X^i d X^j}{\left\|X\right\|^2} \,,
\end{equation}
where $\left\|X\right\|^2$ is defined with respect to the Euclidean norm, i.e., $\left\|X\right\|^2 = \delta_{ij} X^i X^j$ and $\epsilon_{ij}$ is the two-dimensional Levi-Civita symbol, i.e, $\epsilon_{12} = - \epsilon_{21} = 1$.
$I_C$ takes values in $\mathbb{Z}$ and hence cannot change under a continuous deformation of the curve $C$ as long as one does not cross a zero or singularity of $X$.
It is given by the following sum over all zeros and singularities that are (counter-clockwise) encircled by $C$,
\begin{equation}\label{eq:sumindex}
    I_{C}(X) = \sum_i \mathrm{ind}_{x_i} (X) \in \mathbb{Z} \,,
\end{equation}
The index $\mathrm{ind}_{a} (X)$ is defined as the winding number of $X$ around the point $a$.
Concretely, if $D \subset U$ is a small disk with center $a$ such that no other zero or singular point lies within $D$, then $\mathrm{ind}_{a}(X)$ is defined as the degree of the map $X/\left\|X\right\|^2 \colon \partial D \rightarrow S^1$.

It is important to note that the sign of $\mathrm{ind}_{a} (X)$ does not carry any information on whether $a$ is a zero or a pole of $X$.
For example, if we denote coordinates on $\mathbb{R}^2$ by $x$ and $y$, the vector field $X = (x, y)  (x^2 + y^2)^\alpha$ satisfies $\mathrm{ind}_0(X) = 1$ for any real value of $\alpha$.
Similarly, for $X = (x, -y)  (x^2 + y^2)^\alpha$ we have $\mathrm{ind}_0(X) = -1$, while for $X = (x^2, y)  (x^2 + y^2)^\alpha$ we find $\mathrm{ind}_0(X) = 0$.
Therefore, even if $X$ is non-singular, $I_C \neq 0$ is only a sufficient condition for the presence of at least one zero in the interior of the contour $C$, however it is not a necessary condition.

The above results can naturally be  reformulated in terms of a single complex function $f\colon U \subset \mathbb{C} \rightarrow \mathbb{C}$ with
\begin{equation}
f(z = x^1 + i x^2) = X^1(x^1, x^2) + i X^2(x^1, x^2) \,.
\end{equation}
Importantly, for an arbitrary vector field $X$, $f$ is in general not holomorphic, or, in physics notation, $f = f(z, \bar z)$.
A direct calculation shows that the contour integral \eqref{eq:contourindex} becomes
\begin{equation}\label{eq:complexindex}
I_C(X) = \frac{1}{2 \pi i} \oint_C \frac{\partial_z f(z, \bar z) dz + \partial_{\bar z} f(z, \bar z) d\bar z}{f(z, \bar z)}  \,,
\end{equation}
where we made use of the fact that the real part of the logarithm is single-valued, and that $C$ is a closed contour.
If $f$ is meromorphic (i.e.~$\bar \partial_{\bar z} f = 0$),
one can easily show that
$\mathrm{ind}_{a}(X) = n$ if $a$ is a zero with multiplicity $n$, and $\mathrm{ind}_{a}(X) = - m$ if it is a pole of order $m$,
so one recovers Cauchy's argument principle
\begin{equation}\label{eq:cauchy}
\frac{1}{2 \pi i} \oint_C \frac{f'(z)}{f(z)} d z = Z - P \,,
\end{equation}
where $Z$ and $P$ are the numbers of zeros and poles within $C$.

In lieu of giving a full proof of the above results, we just present some basic motivation.
We first notice that the integrand in \eqref{eq:contourindex} satisfies (for $X\neq 0$)  
\begin{equation}\label{eq:closed}
    d \left(\frac{\epsilon_{ij} X^i d X^j}{\left\|X\right\|^2}\right) = 0 \,,
\end{equation}
it is thus closed when seen as a differential form.
Therefore, it must be locally exact.
Indeed, one has
\begin{equation}\label{eq:arctan}
    \frac{\epsilon_{ij} X^i d X^j}{\left\|X\right\|^2} = d \arctan \frac{X^2}{X^1} \,.
\end{equation}
Eq.~\ref{eq:closed} implies that the integral \eqref{eq:contourindex} does not depend on any smooth deformation of the contour $C$ as long as one does not pass through any points where it is ill-defined.
Therefore, we can deform the contour into a sum of arbitrarily small circles, each containing exactly one zero or pole.
We can then use \eqref{eq:arctan} to evaluate the integral for each of these contours, and will eventually recover the sum formula \eqref{eq:sumindex}.
A similar reasoning can be applied to the complex integral \eqref{eq:complexindex}, where the integrand is locally given by $d \log f$, showing that it computes the integrated change of the phase of $f$ along the contour $C$.

There is a direction generalization to the higher-dimensional case.
For $X = (X^1, \dots, X^n)$ a vector field defined on the closure of an open subset  $U \subseteq \mathbb{R}^n$ with finitely many isolated zeros and poles (and none of them on the boundary $\partial U$) the index $ I_{\partial U}$ is defined by the integral
\begin{equation}\label{eq:multidimindex}
    I_{\partial U}(X) = \frac{1}{\mathrm{vol}(S^{n-1})} \oint_{\partial U} \frac{\epsilon_{i_1 \dots i_n} X^{i_1} d X^{i_2} \wedge \ldots \wedge d X^{i_n}}{\left\|X\right\|^n} \,,
\end{equation}
where $\mathrm{vol}(S^{n-1})$ denotes the volume of the $(n-1)$-dimensional unit-sphere.
As above, $I_{\partial U}$ is an integer number and is equal to the sum \eqref{eq:sumindex} of the index of $X$ over all zeros and poles of $X$.
The index $\mathrm{ind}_{a} (X)$ of X at a point $a$ is defined analogously to the two-dimensional case as the degree of the map $X/\left\|X\right\|^2 \colon \partial B_n \rightarrow S^{n-1}$, for $B_n$ a small ball centered around $a$ such that $B_n$ contains no other zeros or singularities of $X$.

Finally, in case we are dealing not only with an open subset of $\mathbb{R}^n$ but with a general
manifold $\cM$,  further complications can arise.
We define the boundary integral $I_{\partial \cM}(X)$ for a vector field $X$ on $\cM$ using \eqref{eq:multidimindex} in local components $X^i$.
It is well-defined as long as the transition functions between different patches on the boundary take values in the orthogonal group $\mathrm{O}(n)$.
We furthermore need to define an auxiliary vector field $\widetilde X$ that is pointing in the outward normal direction along the boundary $\partial \cM$.
$\widetilde X$ is otherwise arbitrary and it is sufficient if it is defined only in a small neighborhood around $\partial \cM$ and not globally on $\cM$. 
One then has
\begin{equation}\label{eq:manifoldindex}
    I_{\partial \cM}(X) - I_{\partial \cM}(\widetilde X) + \chi(\cM)= \sum_i \mathrm{ind}_{x_i} (X)  \,,
\end{equation}
where $\chi(\cM)$ is the Euler characteristic of $\cM$.
If $\cM$ has no boundary or if $I_X = I_{\widetilde X}$, this reduces to the Poincar\'e-Hopf theorem.
Furthermore, this formula is consistent with the previously discussed cases where $\cM$ is a codimension-zero submanifold of $\mathbb{R}^n$ since there $I_{\partial \cM}(\widetilde X) = \chi(\cM)$.

\subsection{Constraints on vacua from boundary integrals}\label{sec:boundarypotentials}

We consider an action of the form
\begin{equation}\label{eq:nonsusyaction}
    S_\phi = \frac12\int d^d x \sqrt{g} \Bigl(R - g_{ij}(\phi) \partial_\mu \phi^i \partial^ \mu \phi^j - 2V(\phi) \Bigr) \,,
\end{equation}
The scalar fields $\phi^i$ take the role of local coordinates on a scalar manifold $\cM$ and $g_{ij}(\phi)$ defines a Riemannian metric on it.
Such an action arises, for example, from compactification of a higher-dimensional supergravity theory so that the $\phi^i$ are deformation parameters or geometric moduli of the internal compactification manifold.
In general, there might also be other fields such as gauge fields and fermions. They are, however, irrelevant for the following discussion. 

Solutions of \eqref{eq:nonsusyaction} that preserve a maximal amount of spacetime symmetries are easily characterized by
\begin{equation}
    \phi^i(x^\mu) = \phi^i_0 = \mathrm{const.} \quad \text{and} \quad \partial_i V \bigr|_{\phi = \phi_0} = 0\,,
\end{equation}
i.e.,~the scalar fields take a constant value so that the potential $V$ sits at a critical point.
Generally, the point $\phi_0 \in \cM$ is not necessarily unique and there might be a left over moduli space of solutions.
Furthermore, the background solution for the spacetime metric $g_{\mu\nu}$ is given by a (locally) maximally symmetric spacetime, i.e., (Anti)-de Sitter or Minkowski space.
The curvature scale or cosmological constant of the background spacetime is set by the vacuum expectation value of the potential $V$,
$
    R_{\mu\nu} = g_{\mu\nu} V|_{\phi = \phi_0} 
$.

We now want to use the techniques outlined in the previous subsection to infer information on the existence of such a solution from the behaviour of the potential $V(\phi)$ on the boundary $\partial \cM$ of the scalar field space.
(Often, $\partial \cM$ should be understood in the asymptotic sense.)
As discussed in the introduction, for concrete examples the potential is often much better understood at its (asymptotic) boundary than at a generic point in the interior of $\cM$, providing physical motivation for this approach.

For this purpose we take the vector field $X$ from Section~\ref{sec:index} to be given by the gradient of the potential,
\begin{equation}
    X_i = \partial_i V \,,
\end{equation}
Let us furthermore assume that $V$ is non-singular anywhere on $\cM$, and that it only has isolated critical points.
We can then use \eqref{eq:multidimindex} to compute the boundary integral $I_{\partial \cM} (\partial V)$, and relate it with \eqref{eq:manifoldindex} to a sum over all critical points of $V$.
Going forward we also take the simplifying assumption that the topology of $\cM$ is sufficiently trivial such that $I_{\partial \cM}(\widetilde X) = \chi(\cM)$ (the generalization is of course straightforward), in order to simplify our expressions.
Then
\begin{equation}
    I_{\partial\cM}(\partial V) = \sum_{\partial V|_{\phi_0} = 0 } \!\!\! \mathrm{ind}_{\phi_0} (\partial V) \neq 0 \,,
\end{equation}
with the sum taken over all critical points of $V$,
becomes a sufficient condition for the existence of at least one critical point of $V$ with non-vanishing index somewhere on $\cM$, and therefore for the existence of a maximally symmetric solution as described above.
However, $I_{\partial\cM}(\partial V)\neq0$ is clearly not necessary for the existence of a critical point.
There can be critical points with vanishing index or two critical points of opposite index can cancel each others contribution.

A further, more physical caveat is that not every critical point of the potential gives rise to an actually stable vacuum.
To be classically\footnote{We do not take possible quantum mechanical tunneling effects into account here.} stable the critical point has to be a minimum of the potential (in the Minkowski case $V(\phi_0) = 0$) or more specifically, all eigenvalues of the Hessian $\nabla^i \partial_j V$ need to be larger than a certain threshold value (given by the Breitenlohner-Freedman bound in the Anti-de Sitter case).

This problem can be resolved if we restrict our analysis to supersymmetric minima of an F-term potential.
In four-dimensional $\cN = 1$ supergravity the scalar field space $\cM$ is a K\"ahler manifold.
Therefore, taking the $\phi^i$ to be complex fields, the moduli space metric can be expressed in terms of a single K\"ahler potential $K(\phi, \bar \phi)$,
\begin{equation}
    g_{i \bar \jmath} = \partial_i \bar \partial_{\bar \jmath} K \,,
\end{equation}
Furthermore, the potential $V$ can be written solely\footnote{We assume the absence of any D-terms.} in terms of $K(\phi, \bar \phi)$ and a holomorphic superpotential $W(\phi)$,
\begin{equation}\label{eq:SUSYpotential}
    V = e^K \left( g^{i \bar \jmath} D_i W \overline D_{\bar \jmath}  \overline W  - 3 \left|W\right|^2 \right) \,,
\end{equation}
where $D_i W$ denotes the K\"ahler covariant derivative
\begin{equation}\label{eq:DW}
    D_i W = \partial_i W + (\partial_i K) W \,.
\end{equation}
This formulation has the convenient side effect that any point where $DW$ vanishes is also a critical point of $V$,
\begin{equation}
    D_i W\bigl|_{\phi = \phi_0} = 0 \qquad \Rightarrow \qquad \partial_i V\bigl|_{\phi = \phi_0} = 0 \,.
\end{equation}
Furthermore, due to the quadratic structure of \eqref{eq:SUSYpotential}, any such point automatically satisfies the above mentioned stability conditions.
This is clear in the Minkowski case ($W|_{\phi = \phi_0} = 0$) and requires a small calculation in the AdS case ($W|_{\phi = \phi_0} \neq 0$).

It is hence natural to compute the boundary integral \eqref{eq:multidimindex} not directly for the gradient of the potential $\partial_i V$ but for $D_i W$, either by interpreting it as a $2 n$ dimensional real vector field, or by using a higher-dimensional generalization of the complex formula \eqref{eq:complexindex}.
Under the same assumptions as above,
\begin{equation}
    I_{\partial \cM} (DW) \neq 0 \,,
\end{equation}
becomes a sufficient condition for the existence of not only a critical point but of a classical stable vacuum with respect to the potential \eqref{eq:SUSYpotential}.
As before, it is not a necessary condition as minima with zero index or cancellations can occur. 
In the following, we illustrate this procedure for a specific example.

\section{M-theory flux vacua}\label{sec:fluxvacua}

In the following, we want to apply the previously outlined technique to classical flux vacua of M-theory compactified on a Calabi-Yau four-fold $X_4$.
Such compactifications are characterized by a choice of four-form flux $G_4$ that satisfies the quantization condition \cite{Witten:1996md}
\begin{equation}\label{eq:fluxquantization}
    G_4 + \frac12 c_2(X_4) \in H^4(X_4, \mathbb{Z}) \,,
\end{equation}
with $c_2(X_4)$ the second Chern class of $X_4$.
Furthermore, $G_4$ is subject to a tadpole cancellation condition
\begin{equation}\label{eq:tadpole}
    \frac12 \int_{X_4} G_4 \wedge G_4 + N_{M2} = \frac{\chi(X_4)}{24} \,,
\end{equation}
where $N_{M2}$ is the number of spacetime filling M2 branes and $\chi(X_4)$ the Euler characteristic of $X_4$.

Supersymmetric vacua are characterized by the self-duality condition
\begin{equation}\label{eq:selfdual}
    G_4 = \star G_4 \,.
\end{equation}
In the complex structure sector, the three-dimensional effective theory resulting from compactification is described in terms of the classical superpotential \cite{Gukov:1999ya}
\begin{equation}\label{eq:GVW}
    W = \int_{X_4} G_4 \wedge \Omega \,,
\end{equation}
for $\Omega$ the holomorphic four-form on $X_4$.
The self-duality requirement \eqref{eq:selfdual} gives rise to the F-term condition
\begin{equation}\label{eq:DW0}
    D_i W = 0 \,,
\end{equation}
where the index $i = 1, \dots, h^{3,1}(X_4)$ runs over all complex structure moduli of $X_4$.
Here, the K\"ahler covariant derivative is defined as in  \eqref{eq:DW} with respect to the K\"ahler potential
\begin{equation}
    K = - \log \int_{X_4} \Omega \wedge \overline \Omega \,.
\end{equation}
Both the K\"ahlerpotential and superpotential depend non-trivially on the complex structure moduli $z^i$ of $X_4$ by means of $\Omega = \Omega(z^i)$.

Additionally, supersymmetric vacua need to satisfy the primitivity condition
\begin{equation}\label{eq:primitiv}
    G_4 \wedge J = 0 \,,
\end{equation}
with $J$ the K\"ahler form of $X_4$.
This is a condition on the K\"ahler moduli sector and
 can be encoded in a second, real superpotential $\widetilde W \sim \int_{X_4} J \wedge J \wedge G_4$ \cite{Haack:2001jz}.
Notably, at the classical level, the holomorphic superpotential $W$ in \eqref{eq:GVW} does not depend on the K\"ahler moduli $t^\alpha$, whereas $\widetilde W$ is independent of the complex structure moduli $z^i$.
Therefore, at this level, the problem of finding supersymmetric minima factorizes. 

Moreover, the contributions of $D_\alpha W$ 
(with respect to the K\"ahler moduli) 
and $|W|^2$ in the three-dimensional analogue of the potential \eqref{eq:SUSYpotential} cancel.
Therefore, if the primitivity condition \eqref{eq:primitiv} holds,
the resulting potential is of no-scale type and reads
\begin{equation}
    V = e^K g^{i\bar \jmath} D_i W \overline D_{\bar \jmath} \overline W \,.
\end{equation}
Hence, every solution to the F-term condition \eqref{eq:DW0} is indeed a minimum of the potential.
For non-primitive fluxes, there are additional terms in $V$.
As in the four-dimensional case, the holomorphic superpotential $W$ receives non-perturbative quantum corrections that also depend on the K\"ahler moduli $t^\alpha$, generating a non-trivial potential for $t^\alpha$ even for primitive fluxes.
However, we will ignore all such corrections here and work purely at the classical level (or in the infinite volume limit), ignoring from now on the K\"ahler moduli sector entirely.

To compute the superpotential \eqref{eq:GVW} we choose an integer basis $A^I$ of four-cycles,
\begin{equation}
    A^I \in H_4(X^4, \mathbb{Z}) \,,\qquad I = 1, \dots, b^4\,.
\end{equation}
By integrating $\Omega$ over $A^I$ we obtain the periods of $\Omega$,
\begin{equation}\label{eq:periods}
    \Pi^I (z^i) = \int_{A^I} \Omega(z^i) \,. 
\end{equation}
The periods are holomorphic functions of the complex structure moduli $z^i$ and uniquely determine the geometry of the complex structure moduli space.
We furthermore define the flux quanta
\begin{equation}
    g^I = \int_{A^I} G_4 \,.
\end{equation}
Depending on $c_2(X_4)$ in \eqref{eq:fluxquantization}, the $g^I$ are integer or half-integer numbers.
Putting both ingredients together, the superpotential \eqref{eq:GVW} becomes a linear combination of the periods,
\begin{equation}
    W(z) = g^I \eta_{IJ} \Pi^J(z) \,,
\end{equation}
where $\eta_{IJ}$ is the inverse of the intersection form
\begin{equation}
\eta^{IJ} = A^I \cdot A^J \,.
\end{equation}

In the remainder of this paper we will try to evaluate the index that we have introduced in Section~\ref{sec:index} for the flux superpotential \eqref{eq:GVW}.
This requires us to define a suitable contour along the boundary of the complex structure moduli space,
as well as to determine the behavior of the periods $\Pi^I(z)$ along this contour.
We will see for a simple example in the next subsection that the periods are singular only at a few (asymptotic) points in moduli space.
If we hence manage to define a contour $C$ that excludes all such singular points, $I_C(DW) \neq 0$ becomes a sufficient condition for an F-term minimum $D_iW = 0$ somewhere in the interior of the complex structure moduli space.

\subsection{The fermat sextic}\label{sec:sextic}

We now proceed to apply these general considerations to an explicit example.
Due to its simplicity we choose the Fermat sextic Calabi-Yau four-fold.%
\footnote{Flux vacua on the Fermat sextic were also discussed in \cite{Braun:2020jrx, Braun:2023edp}.}
It is defined as a hypersurface in complex projective space
\begin{equation}
    \mathbb{P}^6 = (\mathbb{C}^6 \setminus \{0\}) / \mathbb{C}^\star
\end{equation}
where the $\mathbb{C}^\star$ action is defined by 
\begin{equation}
    (x_1, \dots, x_6) \sim (\lambda x_1, \dots, \lambda x_6) \,,\quad \lambda \in \mathbb{C}^\star \,,
\end{equation}
with $x_n$ the coordinates of $\mathbb{C}^6$.
The hypersurfaces defined by the sextic polynomial
\begin{equation}\label{eq:sexticpolynomial}
    P = \sum^6_{n=1} x_n^6 - 6 \psi \prod^6_{n=1} x_n \,,
\end{equation}
are Calabi-Yau for any value of $\psi \in \mathbb{C}$ and non-singular unless $\psi^6 = 1$.

The polynomial \eqref{eq:sexticpolynomial} allows for many other deformations compatible with the Calabi-Yau condition, resulting in a total of $h^{3,1} = 426$ complex structure moduli.
However, only the deformation parametrized by $\psi$ is invariant under the $\mathbb{Z}_6^4$ symmetry group that acts on the projective coordinates $x_n$ by rotation of their complex phases.
It thus singles out a one-dimensional, invariant subspace $\cM_\psi$ of the full complex structure moduli space $\cM_\mathrm{c.s.}(X_4)$ that is spanned by $\psi$.

Similarly, $\mathbb{Z}_6^4$ acts also on the middle cohomology of $X_4$, leaving only a subspace $H^4_\mathrm{sym.} \subset H^4$ invariant.
As recently discussed in \cite{Lust:2022mhk} (see also \cite{Giryavets:2003vd, Denef:2004dm, Cicoli:2013cha, Demirtas:2019sip, Blanco-Pillado:2020wjn, Blanco-Pillado:2020hbw} ),
restricting to fluxes $G^4 \in H^4_\mathrm{sym.}$ provides a consistent truncation of the vacuum conditions $DW=0$ to the invariant subspace $\cM_\psi$, effectively reducing the problem to a complex one-dimensional one.
Equivalently, if one denotes the image of $X_4$ under mirror symmetry by $Y_4$, one can identify $\cM_\psi$ with the complex structure moduli space of $Y_4$, and  $H^4_\mathrm{sym.}$ with $H^4(Y_4)$.
This gives rise to a one-to-one map between $\mathbb{Z}_6^4$ symmetric flux vacua on $X_4$ and unrestricted flux vacua on $Y_4$.

We also need to comment briefly on flux quantization.
The second Chern class $c_2(X_4)$ of the sextic $X_4$ is given by
(see, e.g., \cite{Klemm:2007in})
\begin{equation}
    c_2(X_4) = 15\, J \wedge J \,,
\end{equation}
where $J$ is the generator of $H^2(X_4, \mathbb{Z})$ and has intersection number $\int_{X_4} J \wedge J = 6$.
Since $c_2(X_4)$ is odd, the quantization condition \eqref{eq:fluxquantization} requires us to turn on half-integer $G_4$-flux in the same class.
Notably, this flux violates the primitivity condition \eqref{eq:primitiv}.
The quantization of primitive fluxes in $H^4_\mathrm{prim.}(X_4)$ remains, however, unaffected.%
\footnote{I would like to thank Max Wiesner for valuable discussions on this point.}
The sextic is hence probably only of limited relevance as a physical model, nonetheless it serves as a sensible toy model for our discussion, as long as we restrict our attention to the complex structure sector.

Henceforth we restrict our discussion to the $\mathbb{Z}_6^4$-invariant and primitive part of $H^4(X_4)$ that we denote by $H^4_\mathrm{prim.,\, sym.}(X_4, \mathbb{Z}) \cong H^4_\mathrm{prim.}(Y_4, \mathbb{Z})$.
Fluxes
    $G_4 \in H^4_\mathrm{prim.}(Y_4, \mathbb{Z}) $
are characterized by five integer numbers,
\begin{equation}
     \left( g^I \right) = \left(g^1, \dots, g^5 \right) \in \mathbb{Z}^5 \,.
\end{equation}
The intersection form $\eta_{IJ}$ ($I,J = 1, \dots, 5$) on $H^4_\mathrm{prim.}(Y_4, \mathbb{Z})$ can be inferred from mirror symmetry and is given by
\begin{equation}\label{eq:eta}
     \eta_{IJ}  = \begin{pmatrix}
        0 & 0 & 0 & 0 & 1 \\
        0 & 0 & 0 & 1 & 0 \\
        0 & 0 & 6 & 0 & 0 \\
        0 & 1 & 0 & 0 & 0 \\
        1 & 0 & 0 & 0 & 0 \\
    \end{pmatrix} \,.
\end{equation}
It will be needed for the computation of the superpotential later.
We also note that the Euler characteristic of the sextic is
\begin{equation}
    \chi(X_4) = 2610 \,,
\end{equation}
limiting the maximal possible M2-charge induced by the flux to $Q = \frac12 \int G_4 \wedge G_4 \leq \frac{435}{4}$ via the tadpole cancellation condition \eqref{eq:tadpole}.

To evaluate the GVW superpotential \eqref{eq:GVW} explicitly, we need to know the period integrals \eqref{eq:periods} as functions of the complex structure modulus $\psi$.
They can be determined as solutions of the Picard-Fuchs (PF) differential equation $\mathcal{L}\, \Pi^I = 0$ (see, e.g., \cite{Klemm:2007in}),
\begin{equation}\label{eq:PF}
    \mathcal{L} = \theta^5 - 6 z \prod_{n=1}^6 \left(6 \theta + n\right) \,,
\end{equation}
where $z=(-6 \psi)^{-6}$ and $\theta = z \partial_z$.
As an ordinary differential equation of degree five it allows for five independent solutions.
This nicely reflects the fact that the primitive middle cohomology of $Y_4$ is five-dimensional.
The PF equation \eqref{eq:PF} can be solved by a power series ansatz, also called the Frobenius method.%
\footnote{An explicit implementation of the Frobenius method for the Fermat sextic can be found in the supplementary Mathematica notebook \texttt{frobenius\_sextic.nb}.}
Earlier applications of this method to the mirror sextic can be found in \cite{Grimm:2009ef, CaboBizet:2014ovf, vandeHeisteeg:2024lsa}.
For a detailed discussion of the periods of the Fermat quintic that shares many important features with the sextic case see also \cite{Candelas:1990rm},
as well as \cite{Bastian:2023shf} for a more recent account.
In the following we summarize the most important facts on the construction of the period integrals that will be needed for our later discussion.%
\footnote{I am indebted to
Damian van de Heisteeg and Max Wiesner 
for sharing their knowledge on period integrals and the geometry of complex structure moduli spaces of Calabi-Yau manifolds with me.
I would also like to thank Damian van de Heisteeg for sharing a Mathematica notebook for the computation of three-fold periods with me.}

We start by noting that there is no single power series expansion that converges on the whole moduli space.
Instead, the above PF system admits three different singular points so that the entire complex plane can be covered by three separate series expansions around these points.
A global, integral basis can then be fixed by mirror symmetry and by matching the different series expansions at points where their regions of convergence overlap.

The first singular point is located at $z=0$ (or $\psi = \infty$) and is usually called the large complex structure (LCS) point.
It is the point of maximal unipotent monodromy.
An appropriate ansatz for a series expansion around this point includes $\log$-terms and reads
\begin{equation}\label{eq:LCSexpansion}
    \omega^\text{LCS}_i(x) = \sum_{n = 0}^4 \sum_{m = 0}^\infty a_{i,n,m} \left(\frac{\log x}{2 \pi i}\right)^n x^m \,,
\end{equation}
where $ x = z$.
The coefficients $a_{i,n,m}$ can be fixed iteratively up to five constants of integration by inserting the series ansatz into the PF equation \eqref{eq:PF} and solving it order by order.
For example, it is possible to choose $a_{i,n,m} = 0$ for $n>i$ and $a_{i,i,0}$ ($i=0,\dots,4$) as the constants of integration.%
\footnote{
A convenient redefinition of the coefficients $a_{i,n,m}$ is given by
\begin{equation}
    a_{i,n,m} = \tilde{a}_{i,n,m} \frac{(6m)!}{(m!)^6}  \,,
\end{equation}
as it satisfies $\tilde a_{0,0,m} = \tilde a_{0,0,0}$ for all $m \geq 0$.
This fixes the first period uniquely, up to an overall normalization factor.
The remaining coefficients of the other four periods still take a more complicated form, see for example the appendix of \cite{Candelas:1990rm} for the similar case of the mirror quintic.
}
An integral basis can be fixed by employing the mirror map.
After introducing the change of coordinate
\begin{equation}
    t = \frac{\log x}{2 \pi i} \,,
\end{equation} the leading order period integrals are polynomial in $t$ and can be expressed in terms of topological data of the mirror manifold (see also \cite{vandeHeisteeg:2024lsa} for a complementary approach).
For details of this construction we refer the reader to the literature and only state the result \cite{CaboBizet:2014ovf},
\begin{equation}\begin{aligned}\label{eq:LCSpolynomial}
 \Pi^1 &= 1 + \mathcal{O}(e^{2 \pi i t}) \,, \\
 \Pi^2 &= t + \mathcal{O}(e^{2 \pi i t}) \,, \\
 \Pi^3 &= -\frac12 t^2 + \frac12 t + \frac58 + \mathcal{O}(e^{2 \pi i t}) \,, \\
 \Pi^4 &= - t^3 + \frac{3}{2} t^2 -\frac34 t- \frac{15}{8} + \frac{105 \zeta(3) i}{2 \pi^3} + \mathcal{O}(e^{2 \pi i t}) \,, \\
 \Pi^5 &= \frac{1}{4} t^4 + \frac{15}{8} t^2 - \frac{105 \zeta(3) i}{2 \pi^3} t - \frac{75}{64} +  \mathcal{O}(e^{2 \pi i t}) \,. \\
\end{aligned}\end{equation}
Direct comparison of these expressions with the leading terms in \eqref{eq:LCSexpansion} fixes the remaining five constants $a_{i,i,0}$.
In these coordinates the logarithmic monodromy around $z=0$ corresponds to shifting $t$ as $t \rightarrow t + 1$.
Importantly, the coefficients of the polynomial terms in \eqref{eq:LCSpolynomial} guarantee that this transformation acts by an $\mathrm{SL}(5, \mathbb{Z})$ transformation on the period vector $\Pi^I$.

The second singular point is a conifold point and sits at $z_C = 6^{-6}$ or $\psi^6 = 1$. 
Here, the periods exhibit a monodromy of order 2.
Hence, after writing $x = z_C - z$, a suitable series ansatz reads
\begin{equation}\label{eq:seriesC}
    \omega^\text{C}_i(x) = \sum_{n = 0}^{\infty} b_{i,n}\, x^\frac{n}{2} \,.
\end{equation}
Therefore, around the conifold point, the periods take the general form
\begin{equation}
    \Pi^I(x) = \Pi^I_C(x) + \Pi^I_\mathrm{hol}(x) \,,
\end{equation}
where $\Pi^I_C$ has a branch cut discontinuity ending at $z= z_C$ and $\Pi^I_\mathrm{hol}$ is  holomorphic and does not exhibit a branch cut.
Since all terms with non-fractional powers can be absorbed into the holomorphic part $\Pi^I_\mathrm{hol}$, $\Pi^I_C$ is conveniently expressed in terms of a series expansion of the form \eqref{eq:seriesC} with $b_{i,2n} = 0$, i.e., containing only fractional powers of $(z-z_C)$.
Specifically, one has $\Pi^2_C = \Pi^3_C = \Pi^4_C = 0$ and \cite{Grimm:2009ef}
\begin{equation}\label{eq:conifoldperiods}
    \Pi^1_C =  \Pi^5_C = \frac{1}{\sqrt{3} \pi^2} \left[\left(1-\frac{z}{z_C}\right)^\frac32-\frac{17}{18}\left(1-\frac{z}{z_C}\right)^\frac52 + \dots \right] \,.
\end{equation}
All terms in this expansion except for the universal prefactor are uniquely fixed by the Picard-Fuchs equation \eqref{eq:PF}.
Similarly, the holomorphic parts satisfy $\Pi^1_\mathrm{hol} = -\Pi^5_\mathrm{hol}$ and $\Pi^2_\mathrm{hol}$, $\Pi^3_\mathrm{hol}$, $\Pi^4_\mathrm{hol}$ are independent functions.
As before, one determines all higher order coefficients $b_{i,2n}$ in the series expansion \eqref{eq:seriesC} by iteratively solving the Picard-Fuchs equation.
The lowest order coefficients $\{b_{i,0}, b_{i,2}, b_{i,4}, b_{i,6}\}$ take the role of constants of integration and are fixed by analytic continuation to the large complex structure regime.
We do not state the numerical results here but notice that $\Pi^1 + \Pi^5 = 2\Pi^1_C$ is the only integral linear combination of the periods that vanishes at the conifold point $z=z_C$.
In Appendix~\ref{app:BC} we list a few more, non-holomorphic relationships between the periods.

The third singular point is the Gepner or Landau-Ginzburg point located at $z = \infty$ (corresponding to $\psi = 0$).
It has an order six monodromy, and for $x = 1/z$ we use the series ansatz
\begin{equation}
    \omega^\text{LG}_m(x) = \sum_{n = 1}^{\infty} c_{m,n}\, x^\frac{n}{6} \,.
\end{equation}
A general set of of six linearly independent solutions is given by $c_{m,0} = 0$ and
\begin{equation}\label{eq:LGcoeffs}
    c_{m,n} =  \frac{\alpha^{ (2m+1) n }\Gamma\left(\frac{n}{6}\right)}{\Gamma\left(n\right)\Gamma\left(1-\frac{n}{6}\right)^5} \,, \qquad m = 0, \dots, 4 \,, \quad \text{and} \quad n \geq 1 \,,
\end{equation}
where $\alpha = e^{i \pi /6}$.
The linear combinations corresponding to the integer basis \eqref{eq:LCSpolynomial} are determined by analytic continuation.
We find 
\begin{equation}\label{eq:LGintegerbasis}
    \Pi^I(z) = 
    \frac{1}{36} \begin{pmatrix}
    0 & 0 & 0 & -6 & 0 \\
    1 & 2 & 3 & -2 & -1 \\
    \frac12 & 0 & -\frac32 & -4 & -\frac32 \\
    4 & 8 & 6 & 16 & 2 \\
    0 & 0 & 6 & 0 & 0 
    \end{pmatrix}
    \omega^\text{LG}_m(z^{-1}) \,.
\end{equation}

Around all three points the periods $\Pi^I$ undergo a monodromy.
The transformation matrices can be read off the respective expansions of $\Pi^I$ given above.
For future reference we summarize them here,
\begin{equation}\begin{aligned}
    \label{eq:monodromies}
    M_\mathrm{LCS} =  \!\begin{pmatrix} 
        1 & 0 & 0 & 0 & 0 \\
        1 & 1 & 0 & 0 & 0 \\
        0 & \!\!-1 & 1 & 0 & 0 \\
        \!-4 & \!\!-3 & 6 & 1 & 0 \\
        4 & 7 & \!\!-6 & \!\!-1 & 1 \\
    \end{pmatrix} \!,\;
    M_\mathrm{C} = \!\begin{pmatrix}
        0 & 0 & 0 & 0 & \!\!-1 \\
        0 & 1 & 0 & 0 & 0 \\
         0 & 0 & 1 & 0 & 0 \\
        0 & 0 & 0 & 1 & 0 \\
        \!-1 & 0 & 0 & 0 & 0 \\
    \end{pmatrix} \!,\;
    M_\mathrm{LG} = \!\begin{pmatrix}
        \!-4 & 4 & 0 & \!\!-1 & \!\!-1 \\
        \!-1 & 1 & 0 & 0 & 0 \\
        \!-1 & 1 & 1 & 0 & 0 \\
        7 & \!\!-3 & \!\!-6 & 1 & 0 \\
        \!-1 & 0 & 0 & 0 & 0 \\
    \end{pmatrix}  .
\end{aligned}\end{equation}
They are not fully independent but satisfy the consistency condition 
$
    M_\mathrm{LCS}\, M_\mathrm{C} \, M_\mathrm{LG} = \id
$.
Moreover, they leave the intersection form \eqref{eq:eta} invariant, $M_\mathrm{X}^T \eta M_\mathrm{X} = \eta$.

\subsection{Contour integration for the fermat sextic}\label{sec:sexticintegral}

We continue by using the results from the previous subsection to define and evaluate a contour integral around the boundary of the complex structure moduli space of the mirror sextic.
We choose an integration contour that encircles all three singular points at $z = 0$, $z = 6^{-6}$, and $z = \infty$.
We have to pay particular attention to their monodromies.
The Landau-Ginzburg point has a $\bbZ_6$ monodromy.
It can be unwrapped by going to the covering space and choosing $\psi$ (see the defining equation \eqref{eq:sexticpolynomial}) as a holomorphic coordinate on the moduli space.
For $0 \leq \arg \psi^{-1}  \leq \pi/3$
it is related via
\begin{equation}
    \psi = \frac{1}{6 z^{1/6}} \,,
\end{equation}
to the coordinate $z$ that was used in the previous section.
For other values of $\psi$ we use the monodromy around the Landau-Ginzburg point to map the periods back to the defining sextant,
\begin{equation}\label{eq:LGmonodromy}
\Pi^I( e^{2 \pi i n / 6} \psi ) =  \left(M_\mathrm{LG}^{n} \right)^I{}_J \Pi^J(  \psi ) \,,
\end{equation}
where $M_\mathrm{LG}^{n}$ denotes the $n$-th power of the monodromy matrix $M_\mathrm{LG}$ given in \eqref{eq:monodromies}.

\begin{figure}[htb]
    \centering
    \begin{tikzpicture}[scale=0.85]

    \newcommand{\arrowL}{
        \tikz \draw[very thick, latex-] (-0.1,0) -- (0.1,0);
    }
    \newcommand{\arrowR}{
        \tikz \draw[very thick, -latex] (-0.1,0) -- (0.1,0);
    }

    \filldraw (0,0) circle (2pt);
    \filldraw (2,0) circle (2pt);
    \filldraw (1, 1.73) circle (2pt);
    \filldraw (-1, 1.73) circle (2pt);
    \filldraw (1, -1.73) circle (2pt);
    \filldraw (-1, -1.73) circle (2pt);
    \filldraw (-2, 0) circle (2pt);

    \draw[dashed, thick] (2,0) -- (4.9,0);
    \draw[dashed, thick] (1, 1.73) -- (2.45, 4.24);
    \draw[dashed, thick] (1, -1.73) -- (2.45, -4.24);
    \draw[dashed, thick] (-1, 1.73) -- (-2.45, 4.24);
    \draw[dashed, thick] (-1, -1.73) -- (-2.45, -4.24);
    \draw[dashed, thick] (-2,0) -- (-4.9,0);

    \draw[red, very thick] (2,0.25) arc (90:270:.25);
    \draw[red, very thick] (2,0.25) -- (4,0.25) node[sloped,pos=0.5]{\arrowR};
    \draw[red, very thick, dashed] (4,0.25) -- (5,0.25);
    \draw[red, very thick] (2,-0.25) -- (4,-0.25) node[sloped,pos=0.5]{\arrowL};
    \draw[red, very thick, dashed] (4,-0.25) -- (5,-0.25);

    \draw[red, very thick] (0.78,1.86) arc (150:330:.25);
    \draw[red, very thick] (0.78,1.86) -- (1.84,3.68) node[sloped,pos=0.5]{\arrowR};
    \draw[red, very thick] (1.22,1.61) -- (2.22,3.34) node[sloped,pos=0.5]{\arrowL};
    \draw[red, very thick, dashed] (1.78,3.59) -- (2.28,4.46);
    \draw[red, very thick, dashed] (2.22,3.34) -- (2.71,4.21);

    \draw[red, very thick] (-1.22,1.61) arc (210:390:.25);
    \draw[red, very thick] (-1.22,1.61) -- (-2.27,3.43) node[sloped,pos=0.5]{\arrowL};
    \draw[red, very thick] (-0.78,1.86) -- (-1.84,3.68) node[sloped,pos=0.5]{\arrowR};
    \draw[red, very thick, dashed] (-1.78,3.59) -- (-2.28,4.46);
    \draw[red, very thick, dashed] (-2.22,3.34) -- (-2.71,4.21);

    \draw[red, very thick] (-2,-0.25) arc (270:450:.25);
    \draw[red, very thick] (-2,-0.25) -- (-4.1,-0.25) node[sloped,pos=0.5]{\arrowL};
    \draw[red, very thick] (-2,0.25) -- (-4.1,0.25) node[sloped,pos=0.5]{\arrowR};
    \draw[red, very thick, dashed] (-4,-0.25) -- (-5,-0.25);
    \draw[red, very thick, dashed] (-4,0.25) -- (-5,0.25);

    \draw[red, very thick] (-0.78,-1.86) arc (330:510:.25);
    \draw[red, very thick] (-0.78,-1.86) -- (-1.84,-3.68) node[sloped,pos=0.5]{\arrowL};
    \draw[red, very thick] (-1.22,-1.61) -- (-2.27,-3.43) node[sloped,pos=0.5]{\arrowR};
    \draw[red, very thick, dashed] (-1.78,-3.59) -- (-2.28,-4.46);
    \draw[red, very thick, dashed] (-2.22,-3.34) -- (-2.71,-4.21);

    \draw[red, very thick] (1.22,-1.61) arc (30:210:.25);
    \draw[red, very thick] (1.22,-1.61) -- (2.27,-3.43) node[sloped,pos=0.5]{\arrowR};
    \draw[red, very thick] (0.78,-1.86) -- (1.84,-3.68) node[sloped,pos=0.5]{\arrowL};
    \draw[red, very thick, dashed] (1.78,-3.59) -- (2.28,-4.46);
    \draw[red, very thick, dashed] (2.22,-3.34) -- (2.71,-4.21);

    \draw[red, very thick, dashed] (5,0.25) arc (2.9:57.1:5) node[sloped,pos=0.5]{\arrowL};
    \draw[red, very thick, dashed] (2.28,4.46) arc (62.9:117.1:5) node[sloped,pos=0.5]{\arrowL};
    \draw[red, very thick, dashed] (-2.71,4.21) arc (122.9:177.1:5) node[sloped,pos=0.5]{\arrowL};
    \draw[red, very thick, dashed] (-5,-0.25) arc (182.9:237.1:5) node[sloped,pos=0.5]{\arrowR};
    \draw[red, very thick, dashed] (-2.28,-4.46) arc (242.9:297.1:5) node[sloped,pos=0.5]{\arrowR};
    \draw[red, very thick, dashed] (2.71,-4.21) arc (302.9:357.1:5) node[sloped,pos=0.5]{\arrowR};

    \node at (0.55,0.1) {\scriptsize LG};
    \node at (1.7,0.3) {\scriptsize C};
    \node at (0.55,1.75) {\scriptsize C};
    \node at (-1.4,1.5) {\scriptsize C};
    \node at (-0.55,-1.75) {\scriptsize C};
    \node at (1.4,-1.5) {\scriptsize C};
    \node at (-1.7,-0.3) {\scriptsize C};
    \node at (5.5,2.6) {\scriptsize LCS};

    \draw[thick, ->] (4.6,2.7) -- (5.2,3);
    \node at (5.5,3.1) {$\infty$};
    
    \end{tikzpicture}
    \caption{The complex structure moduli space $\cM_\psi$ of the mirror sextic (the covering space with respect to the LG-monodromy).
    The Landau-Ginzburg (LG) point sits at the origin, there are six conifold (C) points, and the large complex structure (LCS) point is located at complex infinity.
    The conifold points are connected with the large complex structure point by six branch cuts (dashed radial lines).
    In red the integration contour 
    encircling these points as well as the branch cuts.
    It is to be completed around the LCS point at infinity, indicated by the dashed red line.
    }
    \label{fig:modulispace}
\end{figure}
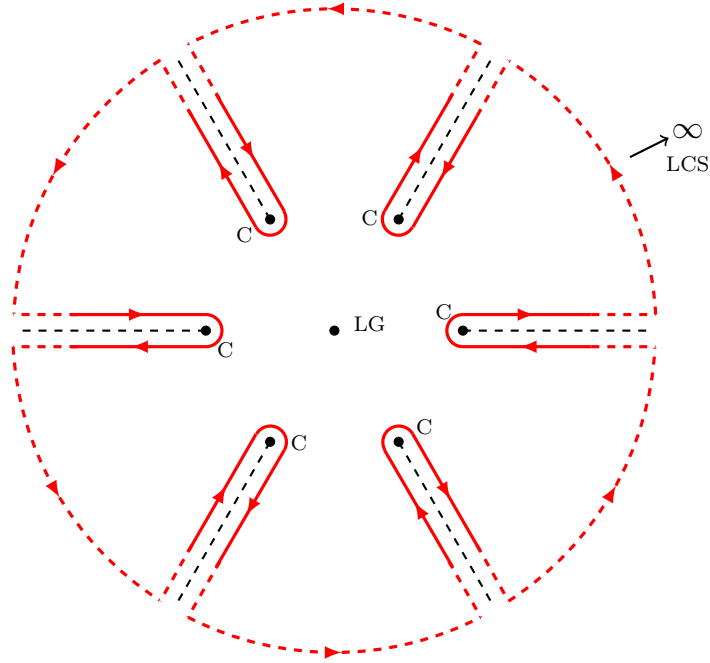

With respect to the $\psi$-coordinate, the Landau-Ginzburg point is at the origin of the complex plane, $\psi = 0$, and has no monodromy. 
Furthermore, there are now six copies of the conifold point at $\psi^6 = 1$ with a $\bbZ_2$ monodromy, and the large complex structure point at complex infinity with a logarithmic monodromy.
Therefore, the conifold points are joined by six branch cuts to the large complex structure point.
The situation is illustrated in Figure~\ref{fig:modulispace}.

Consequently, in order not to cut though the branch cuts, and to stay on one leaf of the moduli space, we choose our integration contour such that it goes not only around the singular points but also follows the six branch cuts on both sides.
These segments connect the contour around the conifold points with the large complex structure points, so that the full contour has only one connected component.
It is depicted in Figure~\ref{fig:modulispace} (in red).

We split up the contour integral into segments that are labelled by their position along the boundary of the complex structure moduli space, and will discuss the computation of each of its pieces subsequently,
\begin{equation}
    I_{\partial \cM_\psi} (DW) = 
    \sum_{i = 0}^5 \left( I^i_\mathrm{LCS} + I^i_\mathrm{b.c.} \right) \,.
\end{equation}
Here $I^i_\mathrm{LCS}$ denotes the contribution from the segment at infinity between the $i$-th and $(i+1)$-th branch cut around the large complex structure point.
In our notation, $I^i_\mathrm{b.c.}$ contains not only the contribution from the branch cut but also from the respective conifold point,
\begin{equation}\label{eq:Ibcsplit}
    I^i_\mathrm{b.c.} = I^i_\mathrm{C} + I^{i,+}_\mathrm{b.c.} + I^{i,-}_\mathrm{b.c.} \,,
\end{equation}
with $I^i_\mathrm{C}$ coming from the $i$-th conifold point, and $I^{i,\pm}_\mathrm{C}$ from the two segments of the contour on either side of the corresponding branch cut.
As we will see later, this simplifies the evaluation of $I^i_\mathrm{b.c.}$ a bit, in particular since generically $I^i_\mathrm{C} =0$.

We begin our discussion with the Landau-Ginzburg point.
It is not directly needed for the computation of the contour integral but helps illustrating a few key points.
We expand the periods with respect to the variable $\psi$,
\begin{equation}
    \Pi^I =  \Pi^I_1 \, \psi + \Pi^I_2 \, \psi^2 + \Pi^I_3 \, \psi^3 + \dots \,,
\end{equation}
where the coefficients $\Pi^I_i$ can be read off from \eqref{eq:LGcoeffs} and \eqref{eq:LGintegerbasis}.
With respect to this expansion we find for the K\"ahler covariant derivative of $\Pi^I$,
\begin{equation}\label{eq:LGDW}
    D_{\psi} \Pi^I = \Pi^I_2 \, \psi + 2 \Pi^I_3 \, \psi^2 - \frac{\Pi^J_2 \eta_{JK} \overline\Pi^K_2}{\Pi^J_1 \eta_{JK} \overline\Pi^K_1} \Pi^I_1 \, |\psi|^2 + \dots \,,
\end{equation}
where we used that
\begin{equation}
    \Pi^I_i \eta_{IJ} \overline\Pi^J_j = 0 \qquad \text{if}\qquad  i \neq j \mod 6 \,.
\end{equation}
Notably, this implies that we always have $D_\psi W = 0$ at $\psi = 0$ independent of the choice of flux vector $g^I$.
This zero of $D_\psi W = 0$ does not correspond to a genuine flux vacuum but would be counted in the index.
To fix this, we choose to compute the index for
\begin{equation}\label{eq:Xchoice}
    X = \psi^{-1} D_\psi W
\end{equation}
instead, which is generically non-vanishing at the origin $\psi = 0$.
To be more specific, $X = 0$ at $\psi = 0$ if
\begin{equation}
    g^I \eta_{IJ} \Pi^I_2 = 0 \,. 
\end{equation}
Using the concrete values for $\Pi^I_2$ that are summarized in Section~\ref{sec:sextic}, we find that this condition is satisfied for
\begin{equation}\label{eq:LGsolutions}
    \left(g^I\right) \in \operatorname{span}_\mathbb{Z} \Bigl\{ (1,0,0,-3,1), (1,1,0,-1,2), (2,1,1,-4,1) \Bigr\} \,.
\end{equation}
The corresponding vacua have index $+1$ or $-1$, depending on the absolute value of the coefficients of the $\psi^2$ or the $|\psi|^2$ term in \eqref{eq:LGDW} when contracted with the flux vector $g_I$.
Notably, there are no flux vectors for which all three terms in \eqref{eq:LGDW} vanish, so  at the Landau-Ginzburg point the index can only take the values $0$ and $\pm 1$.

We continue the discussion with the large complex structure regime.
Here, the contour consists of six individual segments in total, connecting the branch cuts at $|\psi| \rightarrow \infty$.
The periods in the different segments are related by the Landau-Ginzburg monodromy \eqref{eq:LGmonodromy}.
Therefore, the six large complex structure contributions $I^n_\mathrm{LCS}$ to the index are related by transforming the fluxes appropriately,
\begin{equation}\label{eq:LGfluxtransformation}
I^n_\mathrm{LCS}\! \left(g^I\right) = I^0_\mathrm{LCS}\! \left(g^J\! \left(M_\mathrm{LG}^{n}\right)_J{}^I \right) \,,
\end{equation}
where $M_J{}^I = \eta_{IK} M^K{}_L\, \eta^{LJ}$.
It is hence enough to explicitly work out only one of them, say $I^0_\mathrm{LCS}$, as the remaining five parts can be obtained from the same formulae by simply transforming the fluxes.

Restricting now to the sextant $0 \leq \arg \psi^{-1} < \pi / 3$, 
the asymptotic limit of the periods is most easily described in terms of the coordinate
\begin{equation}
t = \frac{3 i}{\pi} \log 6 \psi \,,
\end{equation}
and can be found in \eqref{eq:LCSpolynomial}.
Splitting $t$ into its real and imaginary part,
\begin{equation}
t = x + i y \,,
\end{equation}
the large complex structure limit $|\psi| \rightarrow \infty$ corresponds to sending $y \rightarrow + \infty$.
Furthermore, following the large complex structure contour from the first branch-cut at $\psi \sim 1$ to the previous branch cut at $\psi \sim e^{-2 \pi i/ 6}$ corresponds to sending $x$ from $0$ to $1$.\footnote{Remember that sending $t\rightarrow t+1$ in \eqref{eq:LCSpolynomial} amounts to a counter-clockwise rotation with respect to the coordinate $z$. It hence corresponds to a clockwise rotation of the inverse coordinate $\psi \sim z^{-1/6}$.}
In the following we will demonstrate how to compute the large complex structure contour integral for $D_t W$.
Here, we have to remember that the coordinate change induces a non-trivial Jacobian,
\begin{equation}
D_\psi W \sim \psi^{-1} D_t W \,,
\end{equation}
which we have to take into account when relating it to the index with respect to our previous choice $\psi^{-1} D_\psi W$,
\begin{equation}
I^i_{LCS} (\psi^{-1} D_\psi W) =- \frac13 + I^i_{LCS} (D_t W)  \,.
\end{equation}
As described in Appendix~\ref{app:algorithm}, we can determine $I^i_{LCS} (D_t W)$ without actually computing the integral, simply by tracking the complex phase of $D_t W$ along the contour. 

From \eqref{eq:LCSpolynomial} we find the large-$y$ expansion of $D_t W$ as 
\begin{equation}\begin{aligned}\label{eq:LCSDtW}
D_t W &= - \frac{i}{2} 	g^1 y^3 + \left(g^2 - x g^1\right) y^2 \\
&\;-  \left ( x^3 + \frac{15 x}{4}  -  \frac{525 \zeta(3) i}{16 \pi^3}  \right) g^1  + \left( 3 x^2 - 3 x + \frac34   \right) g^2 + \left(6 x -3 \right) g^3 - g^4  \\
&\; +   \biggl( \frac{i}{32}  \left(16 x^4 + 120 x^2 -75  \right) g^1 - 
     \frac{i}{4} \left(8 x^3 - 12 x^2 + 6 x + 15\right) g^2  +  \frac{105 \zeta(3)}{4 \pi^3}  \left(x g^1 - g^2\right)  \\
     & \qquad \qquad - \frac{3i}{2} \left( 4 x^2- 4 x  -5 \right) g^3 + 2i \left (x g^4 + g^5 \right) 
        \biggr) \frac1y + \mathcal{O}\left(y^{-2}\right) \,.
\end{aligned}\end{equation}
We first notice that if $g^1 \neq 0$ or $g^2 \neq 0$, $D_t W$ does not depend on $x$ in the limit $y \rightarrow \infty$ and is hence constant along the large complex structure contour.
Its complex phase can be read off from the signs of $g^1$ and $g^2$.
In this case, there is no additional contribution to the index so that
\begin{equation}
I^0_\mathrm{LCS}(g^I) = -\frac 13 \,.
\end{equation}
The remaining cases can be inferred from \eqref{eq:LCSDtW} in a similar matter, we relegate their detailed discussion to the Appendix~\ref{app:LCS}.
In particular, it is easy to see that at $y\rightarrow \infty$ and $x=0$ or $x=1$ the leading term in \eqref{eq:LCSDtW} is either real or purely imaginary, and hence
\begin{equation}
I^i_\mathrm{LCS}(g^I) \in  -\frac 13 + \frac{\mathbb{Z}}{4} \,.
\end{equation}

We now go over to the conifold points.
As for the large complex structure contour, it suffices to discuss only one of explicitly.
The remaining ones are related by an appropriate action of the Landau-Ginzburg monodromy $M_\mathrm{LG}$ on the fluxes.
Similarly as before for the Landau-Ginzburg point, we use the expansion of the periods,
\begin{equation}
    \Pi^I = \Pi^I_0 + \Pi^I_{1} (1 - \psi) + \Pi^I_\frac32 (1 - \psi)^\frac32 + \dots \,,
\end{equation}
around the conifold point at $\psi = 1$ to obtain an expansion of $D_\psi \Pi^I$,
\begin{equation}\label{eq:conifoldDPi}
    D_\psi \Pi^I = \frac{\Pi^J_1 \eta_{JK} \overline\Pi^K_0}{\Pi^J_0 \eta_{JK} \overline\Pi^K_0} \Pi^I_0 -\Pi^I_1 -  \frac{3}{2} \Pi^I_\frac32 (1 - \psi)^\frac12 + \dots \,.
\end{equation}
As discussed below \eqref{eq:conifoldperiods}, there is only one linear combination of the periods for which the leading order terms in \eqref{eq:conifoldDPi} vanish, corresponding to the flux vector
\begin{equation}
    g^I \sim \left(1, 0, 0, 0, 1\right) \,.
\end{equation}
This flux corresponds to a flux vacuum exactly at the conifold point.
In this case encircling the conifold point clock-wise gives a contribution of 
\begin{equation}
    I^0_C = - \frac12 \,,
\end{equation}
while for all other fluxes it is identically zero.

It remains to compute the contributions of contours along the branch cuts to the index.
As in the LCS case, the six contour integrals $I^i_\mathrm{b.c.}$ can be related by the same transformation of the fluxes as in \eqref{eq:LGfluxtransformation}.
Moreover, if we split up each contour as in \eqref{eq:Ibcsplit},
we can relate the two segments $I^{i,\pm}_\mathrm{b.c.}$ that follow a branch cut from large complex structure to the conifold point and back by acting on the fluxes with the conifold monodromy ,
\begin{equation}\label{eq:branchcutmonodromy}
I^{i,\pm}_\mathrm{b.c.} \left(g^I\right) = - I^{i,\mp}_\mathrm{b.c.}  \left(g^J\! \left(M_{C}\right)_J{}^I \right) \,.
\end{equation}
Therefore, it is in principle enough to determine only one of them, for example $I^{0,+}_\mathrm{b.c.}$, and to use the monodromies to find the remaining pieces.

Importantly, the branch cuts connect the large complex structure regime with the conifold points.
They are thus not confined to an asymptotic limit where the periods take a particularly simple form but probe the actual interior of the moduli space.
Contrary to the just discussed LCS contours, the contributions of the branch cuts hence cannot be computed  from the leading order terms in a series expansion only but depend on the full periods including all sub-leading terms.
Consequently, in general their purely analytic evaluation appears not to be feasible and one has to rely on numerical methods.

Still, it turns out that a few precise statements can be made.
We refer to Appendix~\ref{app:BC} for more details and only give a brief summary here.
First, the two endpoints of the contour segments that go around the branch cuts are both at asymptotically large complex structure.
Hence, the complex phase of $D_\psi W$ at these points can be read off directly from \eqref{eq:LCSDtW}.
The value of $I^i_\mathrm{b.c.}$ agrees with the difference in phase between the two endpoints, up to a still to be determined integer number.
In the large complex structure limit $y \rightarrow \infty$, the leading term in \eqref{eq:LCSDtW} is either purely real or purely imaginary and therefore
\begin{equation}\label{eq:bcquantization}
    I^i_\mathrm{b.c.} \in \frac{\mathbb{Z}}{4} \,,
\end{equation}
where only the integer part $I^i_\mathrm{b.c.}$ is not yet determined by the large complex structure limit of the periods.

Secondly, as we discuss in the appendix, on the first branch cut (at $\Im \psi = 1$), $\Im D_\psi W$ can be zero only at $\psi = 0$, $\psi \rightarrow \infty$, and at most one other point on the branch cut.
This implies that
\begin{equation}
    \left| I^i_\mathrm{b.c.} \right | \leq 1 \,,
\end{equation}
and restricts, in combination with \eqref{eq:bcquantization}, the possible values of $I^i_\mathrm{b.c.}$ considerably.
Furthermore, $\Im D_\psi W$ has a non-trivial zero on the first branch cut if and only if
\begin{equation}\label{eq:branchcuttest}
    \frac{3 g^3 + g^4}{g^2} > c \,,
\end{equation}
with the value of the the constant $c$ given \eqref{eq:branchcutzerotest}.
In particular, if this inequality is not satisfied, $\Im D_\psi W$ has a definite sign on the branch cut and therefore
\begin{equation}\label{eq:bcbound}
    \left| I^0_\mathrm{b.c.} \right | \leq \frac12 \,.
\end{equation}
In this case, $I^0_\mathrm{b.c.}$ is indeed uniquely determined by the large complex structure limit and can be computed from \eqref{eq:LCSDtW}.
If \eqref{eq:bcbound} is saturated, there is a left over ambiguity that is fixed by the sub-leading behavior of \eqref{eq:LCSDtW}.
The result can be read directly from the signs of $g^1+g^5$, $g^2$, and $3 g^3 + g^4$ and is given in \eqref{eq:nozeroIBC}.

On the other hand, if \eqref{eq:branchcuttest} holds, $I^0_\mathrm{b.c.}$ depends non-trivially on the full periods along the branch cuts.
As discussed in the appendix, it can be computed by determining the phase of $\Re D_\psi W$ at the point where $\Im D_\psi W$ vanishes numerically.
In general, this is a numerically very stable procedure.
Moreover, if the fraction of the fluxes appearing on the left hand side of \eqref{eq:branchcuttest} is large, this point lies far in the large complex structure regime, and the leading order large complex structure periods can be used to approximate $I^0_\mathrm{b.c.}$ with high confidence. 

\subsection{Examples}\label{sec:examples}

We conclude by presenting a few examples that demonstrate the most salient features of our previous discussion.
We leave an exhaustive survey of all flux-vacua on the Fermat sextic for future explorations.%
\footnote{For an exhaustive survey of all flux-vacua of IIB on the full moduli space of the mirror octic using conventional methods see \cite{Plauschinn:2023hjw}.
Flux vacua on toroidal orbifolds were, for example, systematically studied in \cite{Betzler:2019kon, Antoniadis:2024hqd}. 
For recent advances on computer aided searches for flux vacua, see, e.g.,~\cite{Dubey:2023dvu}.
}

\begin{figure}[p]
    \centering
    \begin{tabular}{cc}
    {\small a)}
    \begin{minipage}[c]{0.4\textwidth}
    \includegraphics[width=\textwidth]{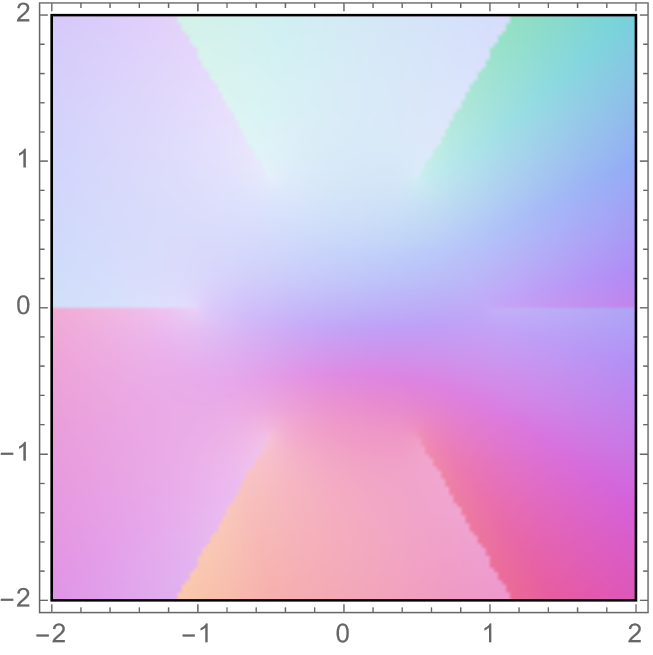}
    \end{minipage}
    &
    {\small b)}
    \begin{minipage}[c]{0.4\textwidth}
    \includegraphics[width=\textwidth]{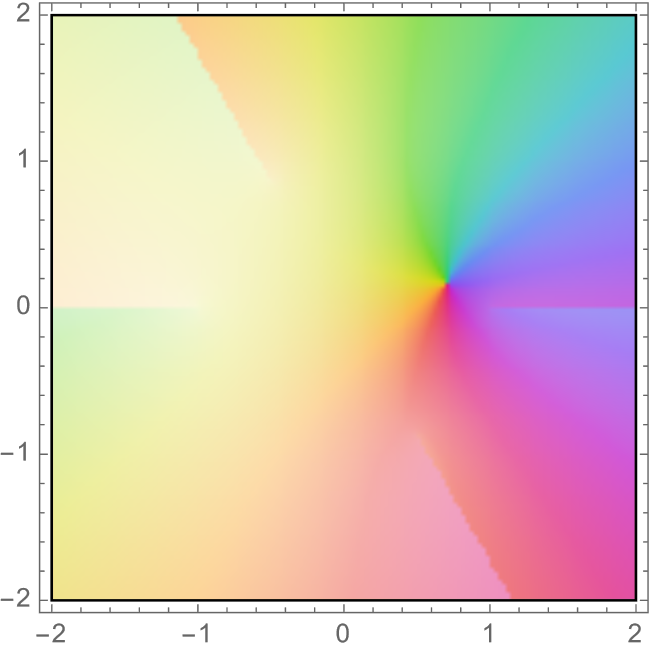}
    \end{minipage}
    \vspace{0.15cm}
    \\

    {\small c)}
    \begin{minipage}[c]{0.4\textwidth}
    \includegraphics[width=\textwidth]{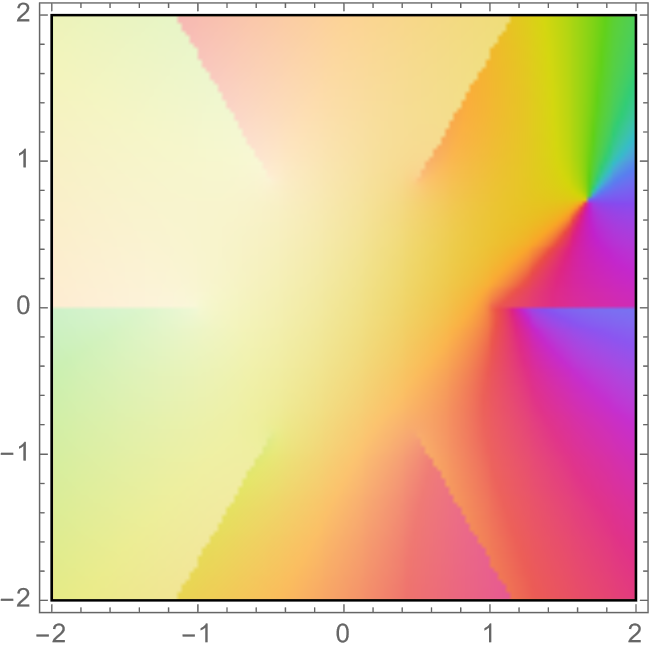}
    \end{minipage}
    &
    {\small d)}
    \begin{minipage}[c]{0.4\textwidth}
    \includegraphics[width=\textwidth]{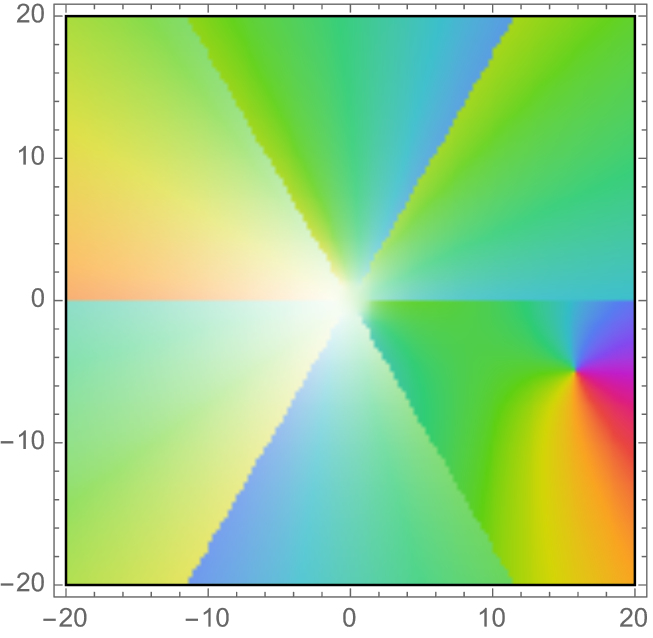}
    \end{minipage}
    \vspace{0.15cm}
    \\

    {\small e)}
    \begin{minipage}[c]{0.4\textwidth}
    \includegraphics[width=\textwidth]{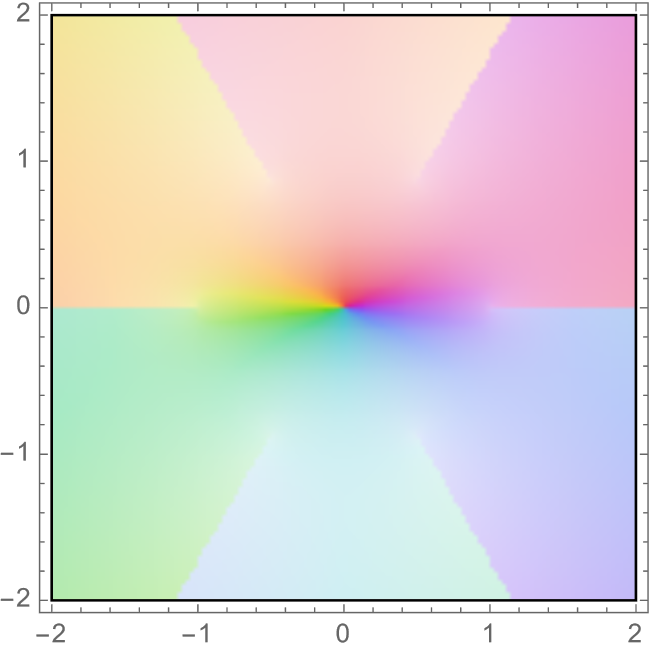}
    \end{minipage}
    &
    {\small f)}
    \begin{minipage}[c]{0.4\textwidth}
    \includegraphics[width=\textwidth]{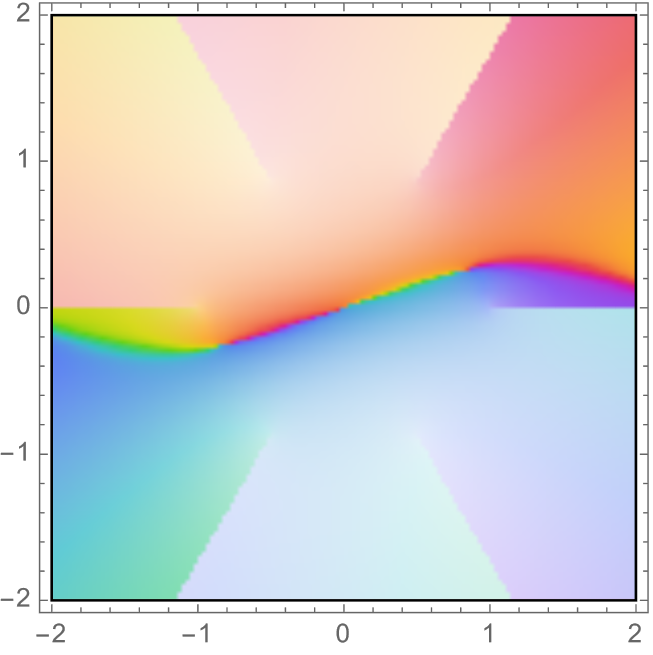}
    \end{minipage}
    \end{tabular}

    
    \caption{Plots of $\psi^{-1} D_\psi W$  on the  $\psi$-plane for different choices of the flux vector $g^I$. Colors encode the complex phase.}
    \label{fig:vacuumplots}
\end{figure}

We illustrate our examples in Figure~\ref{fig:vacuumplots}.
These plots show the complex phase of $\psi^{-1} D_\psi W$ on the complex structure moduli space spanned by $\psi$.
The phase of $\psi^{-1} D_\psi W$ is encoded in a color gradient.
This makes it easy to visually identify vacua by the characteristic phase profile around them.
In order to generate these plots, we computed the periods of the mirror sextic as solutions of the Picard-Fuchs equation \eqref{eq:PF} as high-order series expansions in Mathematica,
following the procedure that is outlined in Section~\ref{sec:sextic}.
This also allows us to verify our findings by searching for solutions of $D_\psi W = 0$ numerically.%
\footnote{The computation of the period integrals is provided in the supplementary notebook \texttt{fermat\_sextic.nb}.
A second notebook, \texttt{flux\_index.nb}, reproduces the examples and plots given in this section and contains routines to compute the index explicitly for arbitrary flux choices, as well as to search for flux vacua numerically.}

\paragraph{a) No solution}
The first flux we discuss is
\begin{equation}
g_a = (1,2,1,1,0) \,,
\end{equation}
and has an M2-tadpole contribution of
    $Q_a = 5$.
    
As discussed around \eqref{eq:LGmonodromy}, we can map the periods on the full $\psi$-plane onto the first sextant by acting with an appropriate power of the Landau-Ginzburg monodromy matrix $M_\mathrm{LG}$ given in \eqref{eq:monodromies}.
This allows us to reduce the computation of the different segments of the contour integral to a repeated evaluation of $I^0_\mathrm{LCS}$ and $I^0_\mathrm{b.c.}$ for all fluxes in the orbit of $g_a$ under the Landau-Ginzburg monodromy.
For this reason we construct the remaining five elements in the $M_\mathrm{LG}$-orbit,
\begin{equation}
    (0, 2, -1, 1, 6)\,, (-6, -4, -3, 13, -5),\, (5, 1, 1, -13, 3),\,   (-3, -2, 0, 2, -3),\, (3, 1, 2, -4, -1) \,. 
\end{equation}
All of them have either $g^1 \neq 0$ or $g^2 \neq 0$. According to our discussion below \eqref{eq:LCSDtW}, this implies that each contribution of the large complex structure part of the contour is trivial, and therefore
\begin{equation}
    \sum_{i=0}^5 I^i_\mathrm{LCS}(g_a) = -2 \,.
\end{equation}
We furthermore observe that none of them satisfies the condition \eqref{eq:branchcuttest}.
Therefore, the branch cut integrals can be obtained by comparing the complex phase of $D_\psi W$ at the two large complex structure end points of the respective branch cut. In case the result is $\pm\frac12$, the remaining ambiguity is fixed by the sign of $g^2 (g^1+g^5)$, see \eqref{eq:nozeroIBC}.
For this example it is always positive.
We find:
\begin{equation}
    I^0_\mathrm{b.c.} = I^1_\mathrm{b.c.} = \frac14 \,,\qquad I^5_\mathrm{b.c.} = 0 \,,\qquad  I^2_\mathrm{b.c.} = I^3_\mathrm{b.c.} = I^4_\mathrm{b.c.} = \frac12 
\end{equation}
and summing everything together we obtain
\begin{equation}
    I(g_a)  = 0 \,.
\end{equation}
Indeed, as one can check via an explicit numerical search, there is no flux vacuum for this choice of flux within our domain.

\paragraph{b) Solution in the Landau-Ginzburg region}
The next flux is very similar and differs from the previous one only in the third flux quantum,
\begin{equation}
    g_b = (1,2,2,1,0) \,,
\end{equation}
resulting in a somewhat higher tadpole contribution,
    $Q_b = 14$.
Again, we compute its orbit under the Landau-Ginzburg monodromy,
\begin{equation}\label{eq:gborbit}
    (0, 2, 0, 7, 0),\, (0, 2, -2, 1, 7),\, (-7, -5, -4, 11, -3), (3, -2, 1, -10, -3), (3, 1, 3, -10, -1) \,,
\end{equation}
and verify that all of them have either $g^1 \neq 0$ or $g^2 \neq 0$ and therefore
\begin{equation}
    \sum I^i_\mathrm{LCS}(g_b) = -2 \,.
\end{equation}
As in the previous example, none of these vectors satisfies the condition \eqref{eq:branchcuttest} so we can go ahead and determine the branch cut contributions directly from the large complex structure phases,
\begin{equation}
    I^0_\mathrm{b.c.} = I^2_\mathrm{b.c.} = \frac14  \,,\qquad  I^1_\mathrm{b.c.} = I^4_\mathrm{b.c.} = I^5_\mathrm{b.c.} = 0 \,,\qquad I^3_\mathrm{b.c.} = \frac12 \,.
\end{equation}
We notice that the first and fourth vector in \eqref{eq:gborbit} satisfy $g^1 + g^5$, so they are invariant under the conifold monodromy and the respective branch cuts disappear.

Summing up all individual contributions we find
\begin{equation}
    I(g_b) = -1 \,.
\end{equation}
Since this is non-zero, we expect a flux-vacuum somewhere.
We can verify this expectation by a numerical search that reveals a zero of $D_\psi W$ at
\begin{equation}
    \psi \approx 0.71 + 0.17 i \,,
\end{equation}
as can be also seen in Figure~\ref{fig:vacuumplots}b.
Notably, this solution lies in the Landau-Ginzburg region, and hence outside of the radius of convergence of the large complex structure periods.
Therefore, this example nicely illustrates that we can predict flux vacua anywhere in the interior of the moduli space using only the periods at large complex structure.

\paragraph{c) Solution in the LCS region}
We increase the value of the third entry of the flux vector again, so our next example reads
\begin{equation}
    g_c = (1,2,3,1,0) \,,
\end{equation}
and has an M2-tadpole contribution of $Q_c = 29$.
We follow our previous two examples and compute the orbit under the Landau-Ginzburg monodromy,
\begin{equation}\begin{gathered}
    (0, 2, 1, 13, -6),\, (6, 8, -1, -11, 19),\, (-19, -11, -9, 35, -9),\,\\ (9, 
-2, 2, -22, -3),\, (3, 1, 4, -16, -1) \,.
\end{gathered}\end{equation}
As before, the large complex structure contribution is directly seen to be trivial,
\begin{equation}
    \sum I^i_\mathrm{LCS}(g_c) = -2 \,.
\end{equation}
However, now $g_b$ and two other elements in the $M_\mathrm{LG}$ orbit actually satisfy \eqref{eq:branchcuttest}.
This means that we cannot directly determine the branch cut integrals from boundary information only but have to follow the numerical procedure that is outlined in Appendix~\ref{app:BC}.
It yields
\begin{equation}
I^0_\mathrm{b.c.} = - I^1_\mathrm{b.c.} = \frac14 \,,\qquad
I^2_\mathrm{b.c.} = - I^3_\mathrm{b.c.} = \frac12 \,,\qquad
I^4_\mathrm{b.c.} = - I^5_\mathrm{b.c.} = \frac12 \,,
\end{equation}
and the total index is given by
\begin{equation}
     I(g_c) = -1 \,.
\end{equation}
As in the previous example, we can numerically confirm the existence of a flux vacuum at
\begin{equation}
    \psi \approx 1.69 + 0.75 i \,,
\end{equation}
see also Figure~\ref{fig:vacuumplots}c.

\paragraph{d) Another solution in the LCS region}
Let us give another example for a solution in the large complex structure region.
Here, we choose the flux
\begin{equation}
    g_d = (0, 1, 0, 20, -4) \,,
\end{equation}
with tadpole
$Q_d = 20$.
This time we refrain from presenting the details of the calculation and only state the final result,
\begin{equation}
    I(g_d) = -1 \,.
\end{equation}
As before, the vacuum can be numerically confirmed and is found at
\begin{equation}
    \psi \approx 16.0 + 5.00 i \,,
\end{equation}
As we can also see from Figure~\ref{fig:vacuumplots}d, it lies far in the large complex structure region.

\paragraph{e) Solution at the Landau-Ginzburg point}
We can also apply the procedure to a solution that sits directly on the Landau-Ginzburg point.
We pick the first vector in \eqref{eq:LGsolutions}, 
\begin{equation}
    g_e = (1,0,0,-3,1) \,,
\end{equation}
with $ Q_e = 1$.
Computing the index via the contour integral gives 
\begin{equation}
    I(g_e) = +1 \,,
\end{equation}
which we have checked to be consistent with \eqref{eq:LGDW}.
Indeed, Figure~\ref{fig:vacuumplots}e shows a zero directly at $\psi = 0$.

\paragraph{f) Multiple solutions}
Another interesting linear combination of the vectors in \eqref{eq:LGsolutions} is given by
\begin{equation}
    g_f = (3, -1, 1, -11, 0) \,,
\end{equation}
with $Q_f = 14$.
Computing the index via the contour integral gives
\begin{equation}
    I(g_f) = +1 \,.
\end{equation}
However, if we now use \eqref{eq:LGDW} to determine the index of the zero at $\psi=0$ directly, we find
\begin{equation}
    \mathrm{ind}_{\psi=0} \left(\psi^{-1} D_\psi W \right) = -1 \,.
\end{equation}
This discrepancy can only be resolved by the presence of at least one other zero somewhere else in the moduli space.
Indeed, we find two more zeros located at
\begin{equation}
    \psi \approx \pm \left( 0.84 + 0.26 i \right) \,,
\end{equation}
and both have index $+1$.
All three zeros are illustrated in Figure~\ref{fig:vacuumplots}f.

\section{Discussion}\label{sec:conclusions}

In this work we suggest to use the winding number of the derivative $DW$ of the superpotential 
to infer information on the existence of supersymmetric vacua. 
We outlined how to compute it as a topological integral around the boundary of the scalar field space, and that this integral equals a sum over all zeros of $DW$ in its interior.
In particular, we showed how to apply this method to classical flux vacua of M-theory described by the GVW superpotential, following the simple example of the sextic Calabi-Yau four-fold.

We demonstrated how to compute the contour integral \eqref{eq:contourindex} on the (asymptotic) boundary of the complex structure moduli space of the mirror sextic, and successfully used the result to predict classical flux vacua for certain explicit choices of fluxes.
The evaluation of the contour integral at the infinite distance limits of field space, in our case the large complex structure point, is relatively straightforward and can be done in closed form.
However, we have found that in order to close the contour, it also needs to probe the interior of moduli space by following the branch cuts that connect different asymptotic limits.
In general, the evaluation of the contour integral along these branch cuts requires the use of numerical methods.
However, we could show that for appropriate choices of the fluxes an analytic computation is still possible.

A natural generalization of this analysis is to extend it to models with multiple complex structure moduli.
This would be particularly interesting in the context of the tadpole conjecture \cite{Bena:2020xrh, Bena:2021wyr}.
It states that in the limit of a large number of complex structure moduli ($h^{3,1} \gg 1$), the M2-charge induced by the flux $G_4$ that stabilizes these moduli is bounded from below as
\begin{equation}\label{eq:tadpoleconjecture}
    \frac12 \int G_4 \wedge G_4 \geq \alpha h^{3,1} \,.
\end{equation}
Depending on the value of the constant $\alpha$, 
such a bound would lead to a potential conflict with the tadpole cancellation condition \eqref{eq:tadpole}, 
This conjecture was tested extensively in the large complex structure regime \cite{Marchesano:2021gyv, Plauschinn:2021hkp, Lust:2021xds, Grimm:2021ckh, Grana:2022dfw, Tsagkaris:2022apo, Coudarchet:2023mmm} and at the Landau-Ginzburg point \cite{Becker:2022hse, Becker:2023rqi}.
However, it is still unclear whether it holds in the interior of complex structure moduli space, where it appears to be possible to violate the bound \eqref{eq:tadpoleconjecture} when going to certain gauge enhancement \cite{Bena:2021wyr, Braun:2023pzd} or highly symmetric \cite{Lust:2022mhk} points in moduli space.
The techniques described here could provide an efficient method 
to go beyond the limitations of previous studies, and 
to probe for the existence of flux vacua in higher dimensional moduli space without restricting to certain asymptotic regimes.
Possible obstacles could, however, arise from the higher-dimensional nature of the boundary of the moduli space itself.

Here, we analyzed the evaluation of the contour integral in the asymptotic limits of moduli space in a rather explicit fashion for one specific example.
However, as for example discussed extensively in \cite{Grimm:2018ohb, Grimm:2018cpv, Grimm:2019wtx, Grimm:2019ixq, Grimm:2020cda, Bastian:2020egp, Grimm:2021ikg, vandeHeisteeg:2022gsp, Grimm:2023lrf}, the asymptotic behavior of the periods can be formalized in terms of asymptotic Hodge theory, revealing various universal properties.
It is hence likely possible to reformulate our analysis in this framework, and to obtain in this way more general results that are not restricted to a specific choice of Calabi-Yau manifold.
However, it is unclear to which extend this approach can be applied to the branch cut contributions that we encountered in our analysis, and that played a crucial role in computing the full contour integral.
Furthermore, it would be interesting to see if the modularity and algebraicity properties of flux vacua that were found in \cite{Kachru:2020sio, Kachru:2020abh, Grimm:2024fip} can be related to our index formulation.

Our analysis was largely motivated by the intuition that effective potentials arising from string compactifications are significantly more easy to compute in asymptotic, infinite distance limits of field space than in its interior.
Ideally, one would use the well-understand asymptotic behavior of a potential in a classical supergravity limit to constrain or predict the existence of critical points or vacua in inaccessible strongly coupled or non-geometric regions.
However, our simple example showed that this might not necessarily always be possible.
In our setup we encountered the difficulty that in order to close the contour  we had to leave the asymptotic boundary of the moduli space and follow the branch cuts towards the interior of the moduli space.
This introduced a dependency on non-trivial information on the potential that could not be inferred from its asymptotic behavior only.
It would be very interesting to see if similar issues also appear in other string theory examples.

It is tempting to reinterpret our findings using mirror symmetry.
Here, we would understand the polynomial terms in the large complex structure expansion of the periods as classical intersection numbers and certain perturbative corrections.
The subleading expontential terms, however, take the role of non-perturbative quantum corrections.
In order to evaluate the contour integral along the branch cuts, one needs to know in principle all of these non-perturbative corrections and cannot rely on the classical or polynomial part only.
However, it is important to stress that this knowledge is enough to predict vacua not only in regions of moduli space where the series expansion of non-perturbative corrections converges but also far outside its radius of convergence in the Landau-Ginzburg phase.
This suggests that it might be indeed possible to predict vacua in a non-geometric or strongly coupled phase of string theory from its classical limit once sufficiently many quantum corrections are also taken into account.

\section*{Acknowledgements}
I would like to thank Paul Balavoine, Markus Dierigl, Mariana Gra\~na,
Damian van de Heisteeg, Dieter L\"ust, Erik Plauschinn, Simon Schreyer, Cumrun Vafa,
and Max Wiesner for very helpful discussions and correspondence.

\appendix

\section{Evaluation of the index}

In this appendix we present more details on the evaluation of the contour integral defining the index.
In Appendix~\ref{app:algorithm} we outline a general algorithm for computing the contour integral without performing an actual integration.
In Appendix~\ref{app:LCS} we discuss the asymptotic contour at large complex structure, and in Appendix~\ref{app:BC} we give a simple recipe to compute the contour integral along the branch cuts.

\subsection{Evaluating the contour integral}\label{app:algorithm}

In the following we present a simple algorithm that can be used to evaluate the contour integral \eqref{eq:contourindex} in the real two-dimensional case.
As already discussed in Section~\ref{sec:index}, the real vector field $X=(X^1, X^2)$ can be encoded in terms of a single complex function
\begin{equation}
    f = X^1 + i X^2 \,,
\end{equation}
with the contour integral given by \eqref{eq:complexindex}.
Importantly, the integrand is locally exact,
\begin{equation}
    I_\mathrm{C}(X) = \frac{1}{2 \pi } \Im \int_C d \log f \,.
\end{equation}
This makes evaluating the integral rather simple with the only subtlety arising from the multivaluedness of the complex logarithm.

Intuitively, $I_\mathrm{C}(X)$ computes the change in the complex phase of $f$ along the contour $C$.
Hence, it can be found by keeping track of the number of different branches of the logarithm it traverses along $C$, or alternatively, how many times $f/|f|$ winds around the origin.
This can be done, for example, by determining all zeros of the imaginary part of $f$ as they indicate the transition onto another branch of the logarithm.
Therefore, $I_\mathrm{C}(X)$ can be computed using the following procedure:
\begin{itemize}
    \item Find all zeros of $X^2 = \Im f$ along the contour $C$, and sort them according to the orientation of the contour with an arbitrarily fixed starting point.
    The first zero should appear twice in the list as both its first and last element.
    \item At each zero, determine the phase of $f$ by evaluating the sign of $X^1 = \Re f$ and denote it by 
    \begin{equation}
        \tilde \phi_i = \frac{1}{2 \pi} \Im \log  f  \in \frac{\mathbb{Z}}{2} \,.
    \end{equation}
    \item Set $\phi_0 = 0$ and then iterate over all points, starting from $i = 1$.
    If 
    \begin{equation}\label{eq:newpointcondition}
         \tilde \phi_{i} - \tilde \phi_{i-1}  \in \mathbb{Z}
    \end{equation}
    set
    \begin{equation}
        \phi_i = \phi_{i-1} \,.
    \end{equation}
    Otherwise, determine the sign $s^i = \pm 1$ of $X^2 = \Im f$ between the two points and set 
    \begin{equation}
        \phi_i = \phi_{i-1} + s_i \left(|\tilde \phi_i| - |\tilde \phi_{i-1}|\right) \,.
    \end{equation}
    \item After the iteration over all points has been completed, the final value for the contour integral can be read off
    \begin{equation}
        I_C(X) = \phi_{N} \,,
    \end{equation}
    with $\phi_{N}$ the last point in the list.
\end{itemize}
Of course, the roles of $X^1$ and $X^2$ in this procedure are completely interchangeable, reflecting the invariance of $I_C(X)$ under the transformation $f \rightarrow \alpha f$ for any $\alpha \in \mathbb{C}  \setminus \{0\}$.

\subsection{Large Complex Structure}\label{app:LCS}

In the large complex structure limit, the index integral can be determined directly from the large $y$ expansion of $D_t W$ in \eqref{eq:LCSDtW}.
As already mentioned there, if $g^1 \neq 0$ or $g^2 \neq 0$, the leading order term in the expansion does not depend on $x$, and is therefore constant along the  contour.
In this case, there is no contribution to the index.
Moreover, the complex phase of $D_t W$ can be read off from the signs of $g^1$ and $g^2$.

On the other hand, if both $g^1 = g^2 = 0$ but $g^3 \neq 0$, eq.~\eqref{eq:LCSDtW} simplifies to
\begin{equation}
D_t W = 3 (2x - 1) g^3 - g^4 - \frac{i}{2 y} \Bigl( 12 (x-1)x g^3  - 15 g^3 - 4 x g^4 - 4 g^5\Bigr)
\end{equation}
and vanishes asymptotically at
\begin{equation}\label{eq:LCSxsolution}
 x = \frac12 + \frac{g^4}{6 g^3} \,.
\end{equation}
If $3 |g^3| < |g^4|$ this value lies outside of the interval $(0, 1)$, 
and hence the phase of $D_t W$ does not vary along the asymptotic contour.
Therefore, also here the contribution to the index is trivial.
The same holds if $g^3 = g^4 = 0$.
In all cases the asymptotic phase of $D_t W$ can be read off the respective leading order terms.

However, if $3 |g^3| \geq |g^4|$ (and $g^3 \neq 0$) there is an actual asymptotic zero within the selected branch of the moduli space.
Consequently, the phase of $D_t W$ is non-constant along the large complex structure contour and gives rise to a non-trivial index.
As long as $3 |g^3| \neq |g^4|$, the asymptotic zero lies within the interior of the interval $(0, 1)$. Therefore, we see 
from the leading order terms that $D_t W$ is purely real at the two endpoints of the contour with opposite signs that are determined by the sign of $g^3$ .
Therefore it contributes with $\pm \tfrac12$ to the index, whereas the sign is determined by the sign of the sub-leading term at $x$ given by \eqref{eq:LCSxsolution}.
If $3 |g^3| \neq |g^4|$ the situation is slightly more subtle, as now $D_t W$ is purely real on one endpoint and purely imaginary at the other endpoint, leading to a contribution of $\pm \tfrac14$ or $\pm \tfrac34$.

In conclusion, we find the following for the large complex structure contribution $I^0_{LCS}$ to the index.
If $g^1 \neq 0$, or $g^2 \neq 0$, or $3 |g^3| < |g^4|$, or $g^3 = g^4 = 0$,
\begin{equation}
I^0_\mathrm{LCS}(g^I) = -\frac 13 \,.
\end{equation}
If $3 |g^3| > |g^4|$,
\begin{equation}
I^0_\mathrm{LCS}(g^I) = -\frac 13 + \frac12 \sign\!\left(9\! \left(g^3\right)^2 + \frac16 \left(g^4\right)^2 + g^3 g^4 + 2 g^3 g^5 \right ) \,,
\end{equation}
if $3 g^3 = - g^4$,
\begin{equation}
I^0_\mathrm{LCS}(g^I) = -\frac 13  + \frac14 \sign \! \left(15\! \left(g^3\right)^2 + 4 g^3 g^5 \right) \,,
\end{equation}
and if $3 g^3 =  g^4$,
\begin{equation}
I^0_\mathrm{LCS}(g^I) = - \frac 13  + \frac14 \sign \! \left(27\! \left(g^3\right )^2 + 4 g^3 g^5 \right) \,.
\end{equation}

\subsection{Branch cuts}\label{app:BC}

Here we give a simple algorithm to evaluate the branch cut integrals $I^i_\mathrm{b.c.}$ exactly.
It requires, in principle, full knowledge of the periods between the large complex structure and the conifold point but no explicit integration.

The contour integral is constructed such that it computes the difference in the complex phase of $D_\psi W$ on the two sides of the branch cut.
As described in \eqref{eq:branchcutmonodromy}, both sides are related by the action of the conifold monodromy.
Hence, we introduce
\begin{equation}\label{eq:Wpm}
    W^+ = W(g^I) \,, \qquad W^- = W\left( g^J\! \left(M_{C}\right)_J{}^I \right) \,,
\end{equation}
so that
\begin{equation}
    I^{i\,\pm}_\mathrm{b.c.} = \pm I^{i,+}_\mathrm{b.c.} \left( X = D W^{\pm} \right) \,.
\end{equation}
This means that we can also compute the contour integral by following the branch cut only once but by adopting the vector field $X$ for which we compute the index accordingly as the ratio of $DW^\pm$, 
\begin{equation}
    \tilde X = \frac{D W^+}{D W^-} \,,
\end{equation}
so that
\begin{equation}
    I^{i}_\mathrm{b.c.}(X) = I^{i,+}_\mathrm{b.c.} (\tilde X  ) \,,
\end{equation}
as this computes exactly the difference in phase between $DW^\pm$.

In the following, we restrict ourselves to the first branch cut at $\Im \psi = 0$ and $\Re \psi \geq 1$ and discuss how to compute $I^{0,\pm}_\mathrm{b.c.}$.
The two sides of the branch cut are related by the action of $M_\mathrm{C}$ given in \eqref{eq:monodromies}.
It acts on the periods by sending
\begin{equation}\label{eq:conifoldaction}
    \Pi^{1,5} \rightarrow - \Pi^{5,1} \,,
\end{equation}
and leaves the remaining three periods $\Pi^{2,3,4}$ invariant. 
We state, without giving a proof, that directly on the branch cut both $\Pi^1$ and $\Pi^5$ are real,
\begin{equation}
    \Im \Pi^1 = \Im \Pi^5 = 0 \,.
\end{equation}
Moreover, the imaginary parts of $\Pi^{2}$ and $\Pi^{3}$ satisfy the linear relation
\begin{equation}
    \Im \Pi^{2} - 2 \Im \Pi^{3} = 0 \,.
\end{equation}
These statements are easily verified for the leading order large complex structure expressions \eqref{eq:LCSpolynomial}, however, they actually hold true at any order.

\begin{figure}[htb]
    \centering
    \includegraphics[width=0.48\textwidth]{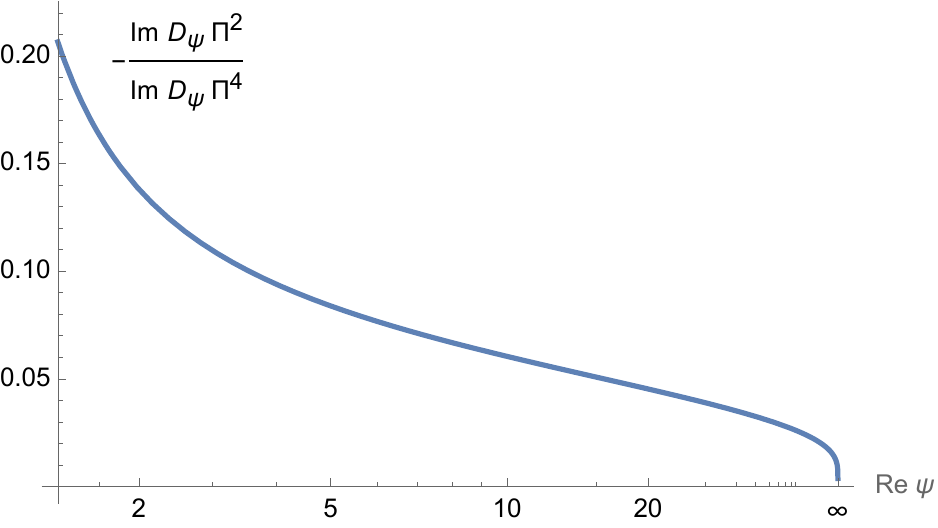}
    \hspace{0.01\textwidth}
    \includegraphics[width=0.48\textwidth]{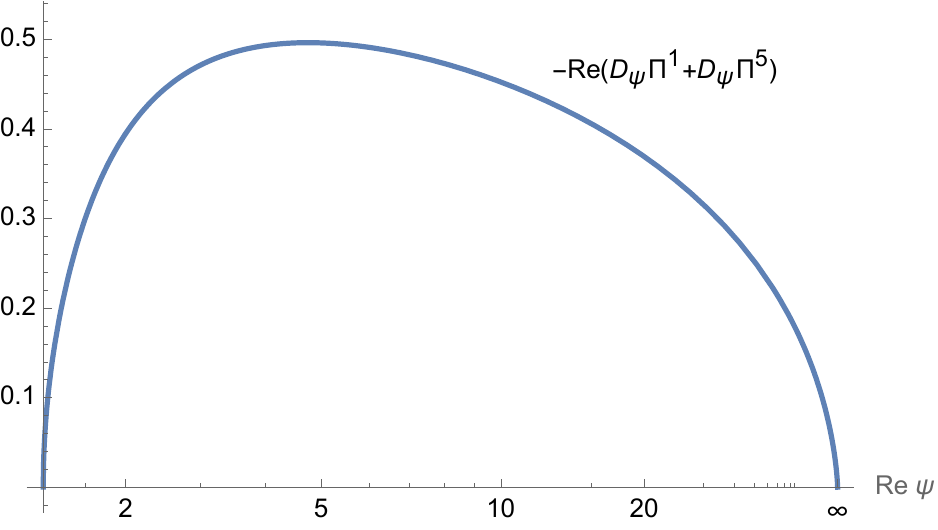}
    \caption{Left: The ratio of the imaginary parts of the periods $D_\psi \Pi^2$ and  $D_\psi \Pi^4$. Right: The real part of the period $- \left(D_\psi \Pi^1 + D_\psi \Pi^5\right)$. Both are evaluated on the first branch cut at $\Im \psi = 0$ and $\Re \psi \geq 1$.}
    \label{fig:periodsplot}
\end{figure}

As a consequence of these observations, the imaginary part of  $D_\psi W$ is invariant under the conifold monodromy and does not differ between the two sides of the branch cut,
\begin{equation}
    \Im D_\psi W^+ = \Im D_\psi W^- \,,
\end{equation}
with $W^{\pm}$ defined in \eqref{eq:Wpm}.
Since $\Im D_\psi W^\pm$ are equal, in the following we will also just write $\Im D_\psi W$ for either of them.
$\Im D_\psi W$ is spanned by only two independent functions, 
\begin{equation}\label{eq:ImDW}
    \Im D_\psi W = \left(3 g^3 + g^4 \right) \Im D_\psi \Pi^2 + g^2 \, \Im D_\psi \Pi^4 \,.
\end{equation}
However, as one can see from Figure~\ref{fig:periodsplot},  the ratio 
\begin{equation}
    - \frac{\Im D_\psi \Pi^2}{\Im D_\psi \Pi^4}
\end{equation}
is a monotonically decreasing function of $\psi \geq 1$, so it takes its maximum at the conifold point $\psi = 1$.
Therefore, $\Im D_\psi W$ has at most one zero at finite $\psi \neq 1$,
and the condition for the existence of such a zero takes the simple form
\begin{equation}\label{eq:branchcutzerotest}
g^2 \neq 0 \qquad \text{and} \qquad \frac{3 g^3 + g^4}{g^2} \geq - \left. \frac{\Im D_\psi \Pi^4}{\Im D_\psi \Pi^2} \right |_{\psi = 1} \approx 4.84 \,.
\end{equation}
Moreover, the larger the ratio on the left hand side, the further the position zero of $\Im D_\psi W$ moves towards the large complex structure point.

Eventually, we are interested in tracking the complex phase of $\tilde X = \frac{D_\psi W^+}{D_\psi W^-}$ along the branch cut.
Its imaginary part satisfies
\begin{equation}\label{eq:ImtildeXdecomp}
     \Im \tilde X \sim \Im  D_\psi W \, \Re\! \left(D_\psi W^+ - D_\psi W^- \right) \,.
\end{equation}
As we have just discovered, $\Im  D_\psi W $ can have at most one zero.
Moreover, from the action \eqref{eq:conifoldaction} of the conifold monodromy on the periods,  we see that
\begin{equation}\label{eq:ReDWpm}
    D_\psi W^+ - D_\psi W^- = \left(g^1 + g^5\right) \left(D_\psi \Pi^1 + D_\psi \Pi^5\right) \,.
\end{equation}
As illustrated in Figure~\ref{fig:periodsplot}, the real part of this linear combination of the periods has a definite sign and is zero only at the two endpoints of the branch cut at $\psi = 1$ and $\psi \rightarrow \infty$.
In combination with our previous observations, this implies that $\Im \tilde X$
also has at most one zero along the branch cut, and that we can use \eqref{eq:branchcutzerotest} to test whether there is such a zero at all.
In particular, this shows that
\begin{equation}
    -1 \leq I^i_\mathrm{b.c.} \leq 1 \,.
\end{equation}
There are two cases, $g^1 + g^5 = 0$ or $g^2 = 3 g^3 + g^4 = 0$, where $\Im \tilde X$ is identically zero along the contour.
We will assume for now that this is not the case and come back to this situation later.
 
To find the actual value of $I^i_\mathrm{b.c.}$ following the general algorithm outlined in Appendix~\ref{app:algorithm}, we need to determine the complex phase 
\begin{equation}\label{eq:tXphase}
    \phi = \frac{1}{2 \pi} \Im \log  \frac{D W^+}{D W^-} 
\end{equation}
of $\tilde X$ at at most three different points.
Firstly, at the conifold point
where $\phi_\mathrm{C} = 0$.
Secondly, at the large complex structure point, where $\phi_\mathrm{LCS}$ can be read off from \eqref{eq:LCSDtW} by setting $x=0$ and sending $y \rightarrow \infty$.
Lastly, if the condition \eqref{eq:branchcutzerotest} is satisfied, also at a third point where $\Im \tilde X$ vanishes.
We denote this third phase by $\phi_0$.

Determining $\phi_0$  requires finding the non-trivial zero of $\Im \tilde X$ numerically, and then probing the sign of $\Re \tilde X$ at this point.
However, if the ratio of the fluxes in \eqref{eq:branchcutzerotest} is large, this point lies well in the complex structure region and $\phi_0$ can be determined from the leading order large complex structure expressions in \eqref{eq:LCSDtW} as higher order terms will be negligible.
Similarly, if the ratio of the fluxes is sufficiently close to the threshold value in \eqref{eq:branchcutzerotest}, the third point lies close to the conifold point and the leading order conifold periods can be used to determine its phase.

In general, \eqref{eq:tXphase} is ambiguous as we have not specified yet on which branch of the complex logarithm we are.
However, this ambiguity is removed by the requirement that the difference in phase between any two subsequent points
($\phi_a$, $\phi_b$) among the three critical points ($\phi_\mathrm{C}$, $\phi_0$, $\phi_\mathrm{LCS}$) is bounded by $|\phi_a - \phi_b| \leq \frac12$.
Otherwise, there would be another zero of $\Im \tilde X$ somewhere between these two points.
If two subsequent points have opposite phases, $|\phi_a - \phi_b| = \frac12$, there is still a left over ambiguity. 
It is fixed by determining the sign of $\Im \tilde X$ along the contour.
Using \eqref{eq:ImtildeXdecomp}, \eqref{eq:ImDW}, and \eqref{eq:ReDWpm} this can be read off from the signs of the fluxes $g^1 + g^5$, $g^2$, and $3 g^3 + g^4$.
Once all such ambiguities are removed, we conclude that
\begin{equation}
    I^0_\mathrm{b.c.}(g^I) = \phi_\mathrm{LCS} \,.
\end{equation}
For example, in the absence of a third phase $\phi_0$, if $|\phi_\mathrm{LCS}| < \frac12$, nothing has to be done and  $I^0_\mathrm{b.c.} = \phi_\mathrm{LCS}$.
However, if $|\phi_\mathrm{LCS}| = \frac12$ we find
\begin{equation}\label{eq:nozeroIBC}
    I^0_\mathrm{b.c.}(g^I) = \frac12\begin{cases}
         \sign g^2 (g^1 + g^5) &\text{if} \quad g^2 \neq 0  \,, \\
        -  \sign \, (3 g^3 + g^4) (g^1 + g^5) &\text{if} \quad g^2 = 0 \,.
    \end{cases}
\end{equation}

As already mentioned above, there are a few singular cases to which the above procedure cannot be applied directly.
Firstly, fluxes that satisfy
\begin{equation}\label{eq:noBC}
    g^1 + g^5 = 0 \,,
\end{equation}
as then $\Im \tilde X = 0$ identically due to \eqref{eq:ImtildeXdecomp} and \eqref{eq:ReDWpm}.
However, in this case the action of the conifold monodromy is trivial, implying the absence of the branch cut.
We can easily deal with this situation by
closing the contour directly at large complex structure and ignoring the respective branch cut segment entirely.
This simply amounts to setting
\begin{equation}
    I^0_\mathrm{b.c.} = 0\,.
\end{equation}
If, in this situation, there is a flux vacuum, $D_\psi W = 0$, directly on the would-be branch cut, it would contribute to the index like any other vacuum anywhere else in the interior of the moduli space.
Such a flux vacuum can, for example, be engineered by imposing the now to be discussed condition \eqref{eq:BCvacuum} in addition to \eqref{eq:noBC}.

The second singular case is characterized by
\begin{equation}\label{eq:BCvacuum}
    g^2 = 3 g^3 + g^4 = 0 \,.
\end{equation}
Here, it follows from \eqref{eq:ImDW} that $\Im D_\psi W = 0$ and thus also $\Im \tilde X = 0$ identically on the branch cut.
This situation requires special treatment if in addition $\Re D_\psi W^\pm =0$ somewhere on the branch cut.
This would imply a flux vacuum with $D_\psi W^{\pm} = 0$ directly on the contour and leads to a potentially ambiguous situation.
We resolve it by infinitesimally deforming the contour on both sides away from the branch cut so that the vacuum lies on the outside of the contour.

The deformed contour can be treated with similar methods as in the general situation that was discussed above.
We only sketch briefly how this works but leave out most of the details:
It turns out that it is easiest to analyse the two segments $I^{0,\pm}_\mathrm{b.c.}$ of the deformed contour separately.
Assuming that $g^1 + g^5 \neq 0$, $\Im D_\psi W^{\pm}$ has a definite sign on each of the two segment of the contour.
This implies that $I^{0,\pm}_\mathrm{b.c.} \in \{0, \pm \frac12\}$.
Moreover, it is enough to determine the sign of $\Re D_\psi W^{\pm}$ at $\psi = 0$ and $\psi \rightarrow \infty$, and the sign of $\Im D_\psi W^{\pm}$ at $\psi \rightarrow \infty$ to determine $I^{0,\pm}_\mathrm{b.c.}$ exactly.
The latter can be inferred by taking the derivative of $\psi^{-2} D_t W$ in \eqref{eq:LCSDtW} with respect to $x$.
A special sub-case is $g^I \sim (1,0,0,0,1)$.
As we saw above, this flux leads to a vacuum directly at the conifold point $\psi = 0$,
and the corresponding deformed contour yields $I^{0}_\mathrm{b.c.} = - \frac12$.

The procedure that we described above yields an algorithm that determines $I^0_\mathrm{b.c.}$ exactly.
However, in general it requires evaluating the periods along the branch cut in order to solve $\Im \tilde X = 0$ and to determine the sign of $\Re \tilde X$ at this point.
In principle, this necessitates knowledge of the full periods along the branch cut and not only of their asymptotic boundary behavior at $\psi = 1$ or $\psi \rightarrow \infty$.
Still, this procedure is numerically very stable as the result is already fixed up to an integer number by the phase of $\tilde X$ at the large complex structure point.
Numerical issues might only arise if the zeros of $\Im \tilde X$ and $\Re \tilde X$ lie very close together on the contour.
Therefore, in practice not many terms in the expansion of the periods around either the conifold or the large complex structure point are needed to obtain an accurate result.
Moreover, the ratio of the fluxes in \eqref{eq:branchcutzerotest} can be used as a proxy to determine if higher order terms have to be taken into account at all.
Finally, if this ratio is below a threshold value, $\Im \tilde X$ is guaranteed to have a definite sign, and $I^0_\mathrm{b.c.}$ is uniquely fixed by boundary information only.


\hskip 1cm


\bibliographystyle{bibstyle}
\bibliography{references}

\end{document}